\documentclass[12pt]{article}

\newenvironment{wileykeywords}{\textsf{Keywords:}\hspace{\stretch{1}}}{\hspace{\stretch{1}}\rule{1ex}{1ex}}

\usepackage{amsmath,amssymb}
\usepackage{graphicx}
\usepackage{color}
\usepackage{dcolumn}
\usepackage{bm}
\usepackage[numbers,super,comma,sort&compress]{natbib}

\definecolor{background-color}{gray}{0.98}

\title{Topological analysis of chemical bonding in the layered FePSe$_3$ upon pressure-induced phase transitions}

\author{Robert A. Evarestov\thanks{Department of Quantum Chemistry, Saint Petersburg State University, 7/9 Universitetskaya Naberezhnaya, St. Petersburg 199034, Russian Federation; E-mail: r.evarestov@spbu.ru}
 and 
Alexei Kuzmin\thanks{Institute of Solid State Physics, University of Latvia, Kengaraga street 8, LV-1063, Riga, Latvia; E-mail: a.kuzmin@cfi.lu.lv}}

\begin{document}

\maketitle

\begin{abstract}
Two pressure-induced phase transitions have been theoretically studied in the layered iron phosphorus triselenide (FePSe$_3$). Topological analysis of chemical bonding in FePSe$_3$ has been performed based on the results of first-principles calculations within the periodic linear combination of atomic orbitals (LCAO) method with hybrid Hartree-Fock-DFT B3LYP functional.  
The first transition at about 6 GPa is accompanied by the symmetry change from $R\bar{3}$ to $C2/m$, whereas the semiconductor-to-metal transition (SMT) occurs at about 13 GPa leading to the symmetry change from $C2/m$ to $P\bar{3}1m$. 
We found that the collapse of the band gap at about 13 GPa occurs due to changes in the electronic structure of FePSe$_3$ induced by relative displacements of phosphorus or selenium atoms along the $c$-axis direction under pressure.
The results of the topological analysis of the electron density and its Laplacian  
demonstrate that the pressure changes not only the interatomic distances but also the bond nature between the intralayer and interlayer phosphorus atoms.
The interlayer P--P interactions are absent in two non-metallic FePSe$_3$ phases while 
after SMT the intralayer  P--P interactions weaken and the interlayer P--P interactions appear.

\end{abstract}

\begin{wileykeywords}
FePSe$_3$, layered compound, topological analysis, high pressure, semiconductor-to-metal transition, first-principles calculations.
\end{wileykeywords}

\clearpage


\begin{figure}[h]
\centering
\colorbox{background-color}{
\fbox{
\begin{minipage}{1.0\textwidth}
\includegraphics[width=50mm,height=50mm]{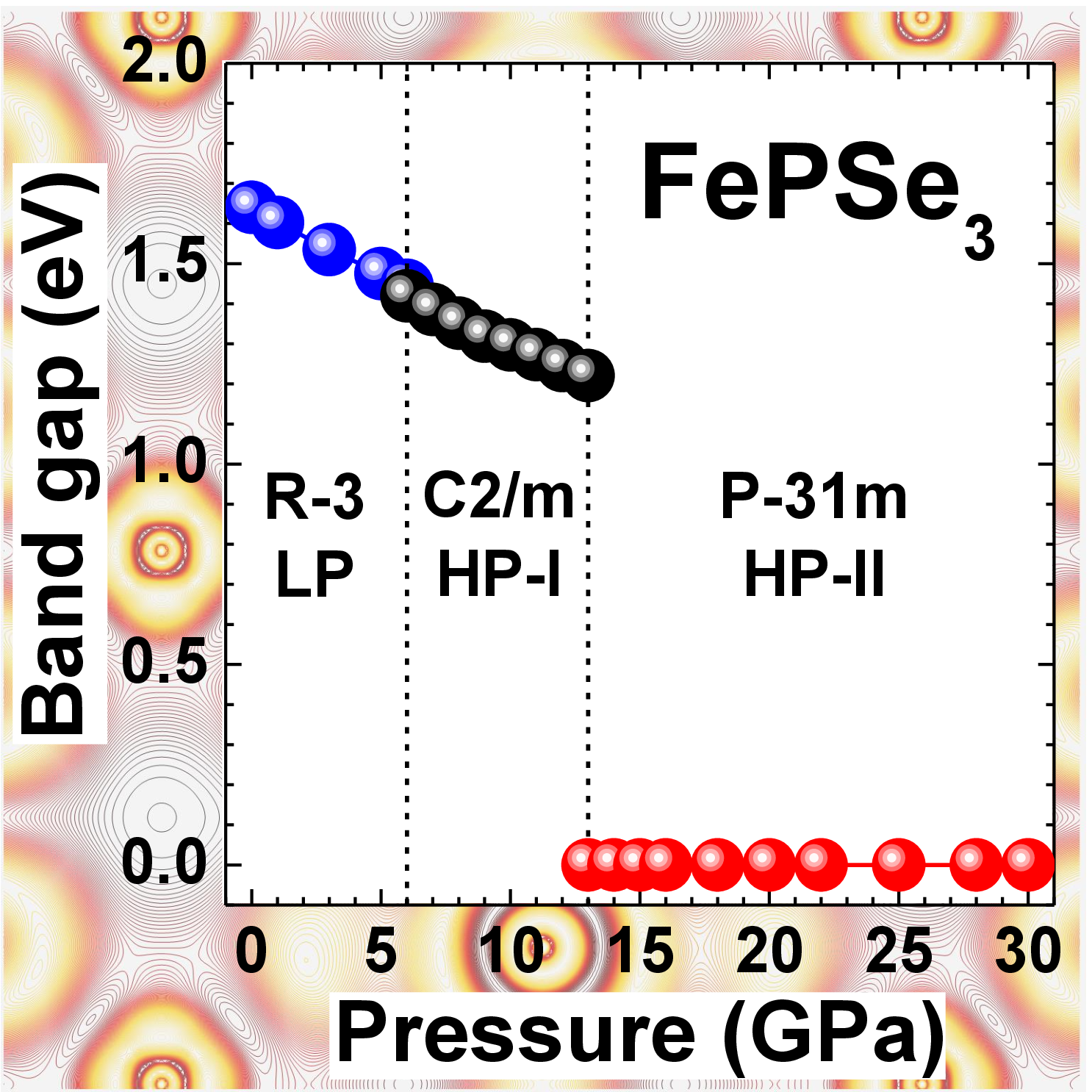} 
\\
Pressure-induced phase transitions in iron phosphorus triselenide (FePSe$_3$) were studied using first-principles calculations and the topological analysis of the electron density and its  Laplacian. The calculations predict the semiconductor-to-metal transition (SMT) at $\sim$13 GPa. The origin of the SMT is attributed to the pressure-induced changes in the FePSe$_3$ electronic structure caused by the relative displacement of phosphorus or selenium atoms along the $c$-axis direction.
The pressure strongly affects both the interatomic distances and the bond nature between the intralayer and interlayer phosphorus atoms. 
\end{minipage}
}}
\end{figure}

  \makeatletter
  \renewcommand\@biblabel[1]{#1.}
  \makeatother

\bibliographystyle{apsrev}

\renewcommand{\baselinestretch}{1.5}
\normalsize

\clearpage

\section{\sffamily \Large INTRODUCTION} 

Topological analysis of the electron density distribution in materials is an efficient method of obtaining information about the existence and nature of chemical bonds.\cite{Bader1994,LEPETIT2017} Due to the progress in quantum chemistry calculations, the method has gained popularity and finds numerous applications in studies of organic and inorganic compounds.\cite{Vyboishchikov1997,Gatti2005,Johnson2010,Matta2011,Krawczuk2014,Marana2016,Dittrich2017,Macchi2017,Varadwaj2019,Titah2019,LACERDA2019,Korabelnikov2019,Tolborg2019}
In particular, the method helps to identify changes in chemical bonds and mechanisms of phase transitions under high pressure.\cite{Parisi2012,Merli2013,Salvado2013,Casati2016,Pereira2016,Merli2018,Belarouci2018,Schwarz2019,Meyer2019,Parisi2019,Gajda2020}
At the same time, the use of topological analysis for the understanding of bonding in layered compounds is not numerous.
Recently it was successfully employed to evidence peculiarities of chemical bonding in layered TiS$_2$ crystals using the analysis of the electron density, obtained from synchrotron X-ray diffraction data and density functional theory calculations.\cite{Kasai2018} 
The method was also used to study the evolution of chemical bonding during solid-state
D$_{\rm 3h}$ $\rightarrow$ D$_{\rm 3d}$ reaction in monolayered TiS$_2$.\cite{Ryzhikov2014}

In this study, we apply the topological analysis to probe in detail both intralayer and interlayer interactions between atoms being responsible for the recently found\cite{Wang2018} pressure-induced semiconductor-to-metal (SMT) transition in iron phosphorus triselenide (FePSe$_3$). 

Bulk FePSe$_3$ is a layered magnetic semiconductor with the narrow optical indirect band gap of 1.3 eV\cite{Brec1979} and possesses intrinsic antiferromagnetism below the N\'{e}el temperature of about 112-123 K.\cite{Taylor1974,WIEDENMANN1981,LeFlem1982}  The magnetism is caused by a 
high-spin state of Fe$^{2+}$ ions (d$^6$) ordered on a honeycomb lattice with the total spin projection  $S = 2$ and the magnetic moments 
$m_0 = 4.9\mu_B$.\cite{WIEDENMANN1981,LeFlem1982} The weak van der Waals (vdW) interactions between neighbouring layers in FePSe$_3$ make its structure and properties strongly  pressure dependent.
The application of high pressure induces the high-to-low spin-state transition
of Fe$^{2+}$ ions, thus producing a nonmagnetic FePSe$_3$ phase through the pressure-induced spin-quenching ($S=0$).\cite{Wang2018,Zheng2019}  
Moreover, it was discovered recently that abrupt spin-crossover is accompanied by SMT and the appearance of a superconducting state
with a transition temperature $T_C \sim 2.5$~K at 9 GPa and the maximum $T_C \sim 5.5$~K at about 30 GPa.\cite{Wang2018} 

Besides, FePSe$_3$ demonstrates enhanced electrocatalytic activity when it is exfoliated into single and few-layer sheets\cite{Gusmao2017,Mukherjee2018,Barua2019,Hao2020} or produced as nanoparticles.\cite{Yu2019}  The layered structure 
of FePSe$_3$ makes it also the material under consideration for cathodes in lithium batteries\cite{Brec1979,Brec1986} and for hydrogen storage.\cite{Ismail2010}

At ambient, low pressure (LP), FePSe$_3$ exists in the trigonal (space group no. 148, $R\bar{3}$) phase.\cite{WIEDENMANN1981} Its  crystallographic structure (Fig.\ \ref{fig1}) is composed of 2D layers, which are parallel to the $ab$-plane and separated by the vdW gaps along the $c$-axis.\cite{WIEDENMANN1981} Each layer consists of Fe atoms octahedrally coordinated by six Se atoms and P atoms tetrahedrally coordinated by three Se atoms and one P atom, forming a [P$_2$Se$_6$]$^{4-}$ unit.  The  shortest distance between the adjacent Se layers (the vdW gap) is  about  3.2~\AA. 
There are two FePSe$_3$ formula units in the primitive unit cell but six formula units in the conventional unit cell
with the atoms occupying the following Wyckoff positions: Fe 6c(0, 0, z), P 6c(0, 0, z) and Se 18f(x, y, z). 

Upon application of pressure, FePSe$_3$ transforms into the monoclinic HP-I phase (space group no. 12, $C2/m$)
with two FePSe$_3$ formula units in the primitive unit cell but four formula  units in the conventional unit cell (Fig.\ \ref{fig1}). The transition pressure was predicted theoretically to be about 6 GPa.\cite{Zheng2019}
The adjacent layers are separated by the narrower vdW gap of about 2.95~\AA, besides, layer sliding occurs in the $ab$-plane placing the P--P atom pairs in the neighbouring layers on top of each other along the $c$-axis. 
The atoms occupy the following Wyckoff positions: Fe 4g(0, y, 0), P 4i(x, 0, z), Se1 4i(x, 0, z) and Se2 8j(x, y, z). 

In the high-pressure (HP-II) phase (Fig.\ \ref{fig1}), observed experimentally above 8 GPa,\cite{Wang2018} FePSe$_3$ crystal belongs to the space group no. 162, $P\bar{3}1m$, with the hexagonal lattice and two formula units in the primitive unit cell. The atoms occupy the following Wyckoff positions: Fe 2c(1/3, 2/3, 0), P 2e(0, 0, z) and Se 6k(x, 0, z).  The transition to the HP-II phase is accompanied by abrupt spin-crossover transition and semiconductor-to-metal transition.\cite{Wang2018} The latter is mainly determined by the in-plane metallization.\cite{Wang2018} Note also that the resistivity of FePSe$_3$ changes stronger with pressure than with temperature.\cite{Wang2018}

The mechanism of the pressure-driven SMT transition in FePSe$_3$ was studied\cite{Zheng2019} using first-principles calculations based on the plane-wave density functional theory (DFT) in the range from 0 to 35 GPa.  The LP-to-HP-I phase transition was found to occur at about 6 GPa, whereas 
the second transition to the HP-II phase was predicted at 15 GPa.\cite{Zheng2019}
The antiferromagnetic ordering was reproduced in the LP and HP-I phases in agreement with the experiment. The band gap of about 1.24 eV was found in the LP phase. It decreases to 0.72 eV at 10 GPa in the HP-I phase, and, finally, drops to zero in the metallic HP-II phase.\cite{Zheng2019}   

In our preceding study,\cite{Evarestov2020} we considered the  pressure-induced SMT in the crystalline iron thiophosphate (FePS$_3$) using first-principles DFT calculations, which confirmed the experimentally found SMT at about 15 GPa. Moreover, we found from the detailed analysis of the projected density of states that the 3p states of phosphorus atoms contribute significantly at the bottom of the conduction band,
and the effect can be tuned by the relative displacement of phosphorus or sulfur atoms along the $c$-axis. Therefore, the collapse of the band gap occurs due to changes in the electronic structure of FePS$_3$ induced by relative displacements of phosphorus or sulfur atoms under pressure. 

In the present study of SMT in FePSe$_3$, we made the same conclusion based not only on the change of the relative displacements of phosphorus or selenium atoms but also on the topological analysis of the pressure dependence of the P--P and Se--Se chemical bonds.
The results of the topological analysis  demonstrate that  the new bonds appear between phosphorous atoms located in the neighbouring layers in the HP-II phase while they are absent in the other two phases (LP and HP-I).

\section{\sffamily \Large METHODOLOGY}

In the present study, the pressure-induced phase transitions in FePSe$_3$ have been investigated using the first-principle linear combination of atomic orbitals (LCAO) calculations using the CRYSTAL17 code.\cite{crystal17}
The basis sets for Fe, P, and Se atoms have been chosen in the form of all-electron triple-zeta valence (TZV) basis sets augmented by one set of polarization functions (pob-TZVP).\cite{Oliveira2019}

The evaluation of the Coulomb series and the exchange series has been done with the accuracy controlled by a set of tolerances, which were taken to be (10$^{-8}$, 10$^{-8}$, 10$^{-8}$, 10$^{-8}$, 10$^{-16}$). The integration in the Brillouin zone has been performed 
using the Monkhorst-Pack scheme \cite{Monkhorst1976} for an 8$\times$8$\times$8 $\textbf{k}$-point mesh.  The SCF calculations were performed using Becke's 3-parameter functional (B3LYP-13\%) \cite{B3LYP}  with a 10$^{-10}$ tolerance on change in the total energy as in our previous study on FePS$_3$.\cite{Evarestov2020}
The percentage (13\%) defines the Hartree-Fock admixture in the exchange part of DFT functional. All calculations were performed using a restricted closed-shell hamiltonian, i.e., for non-magnetic structures. 
We believe that such approximation is consistent with the experimental temperature dependence of the electrical resistance  (see Fig.~3 in Ref.~\cite{Wang2018}), which demonstrates the SMT in FePSe$_3$ in a wide range of temperatures up to 300 K, i.e., far above its N\'eel temperature of $T_N$ = 119 K.\cite{WIEDENMANN1981} 

The lattice parameters and atomic fractional coordinates were optimized for each selected pressure in the range of 0--30 GPa for three phases (Fig.\ \ref{fig2}):
low-pressure (LP) trigonal (space group  $R\bar{3}$) phase with rhombohedral lattice, high-pressure (HP-I) monoclinic (space group $C2/m$) phase, and high-pressure (HP-II) trigonal (space group $P\bar{3}1m$) phase with a hexagonal lattice. The starting structural parameters were taken from the experimental data for the LP FePSe$_3$ 
phase\cite{WIEDENMANN1981} and the related HP-I and HP-II FePS$_3$ phases.\cite{Haines2018}
The structure optimization at the required pressure was performed using the approach developed in Ref.\ \cite{Jackson2015}.
The calculated lattice parameters ($a$, $b$, $c$, $\beta$), atomic fractional coordinates  ($x$, $y$, $z$), the values of the band gap $E_g$ and the atomic charges are reported in Table\ \ref{table1}. The data for the FePSe$_3$ $R\bar{3}$ phase are reported in the hexagonal setting. 
The atomic charges were obtained using two methods -- from the charge density distributions using the Bader procedure\cite{Bader1994} and from the Mulliken population analysis.\cite{Mulliken1955} 

Calculated  band structures and total/projected density of states (DOS) for the LP, HP-I and HP-II FePSe$_3$ phases are shown in Figs.\ \ref{fig3} and \ref{fig4}, respectively.
The high-symmetry points in the Brillouin zone for the three phases were selected following their definition on the Bilbao Crystallographic Server.\cite{Aroyo2014}
Note that in the rhombohedral space group $R\bar{3}$, belonging to the trigonal crystal system, two possible shapes of the Brillouin zone (and, respectively, sets of the high-symmetry points) are possible depending on the 
ratio between $a$ and $c$ lattice parameters.\cite{Aroyo2014} 
In our case $\sqrt{3}a < \sqrt{2}c$ (Table\ \ref{table1}) therefore the Brillouin zone has the topology of the truncated octahedron.   

The phonon frequencies were computed at the center of the Brillouin zone (the $\Gamma$-point)  within the harmonic approximation using the direct (frozen-phonon) method \cite{crystal17,Pascale2004} for each FePSe$_3$ phase. The primitive cell of FePSe$_3$ in all phases includes 10 atoms (2 Fe, 2 P, and 6 Se), so that 30 phonon modes are expected and are classified as the three acoustic (A) modes with zero frequency, the Raman-active (R), infrared-active (IR) and silent (S) modes.
The phonon frequencies calculated for three phases at $P$=0, 10, and 20 GPa  are reported in Table\ \ref{table2}.
Topological analysis of the electron density, according to the Quantum Theory of Atoms in Molecules,\cite{Bader1994} was performed using the TOPOND code\cite{Gatti1994} in its parallel version,\cite{Casassa2015} which is incorporated in the CRYSTAL17 code.\cite{crystal17}
Additionally, the calculations were also performed using the PBE0 \cite{PBE0} and M06 \cite{M06}  functionals to check the sensitivity of the results of the topological analysis to the functional type.

Finally, we have performed the calculations of the electronic structure of FePSe$_3$ for several model structures with P atoms displaced along the $c$-axis. The crystal structures were fixed at the ones optimized for LP ($R\bar{3}$, 0 GPa), HP-I ($C2/m$, 10 GPa), and HP-II ($P\bar{3}1m$, 20 GPa) phases, while the displacement $\Delta z$(P)  of the phosphorus atoms   was varying between $-$0.1 \AA\ and 0.6 \AA. The obtained variations of the band gap  $E_g$ are shown  in Fig.\ \ref{fig5}. 
Total and projected density of states (DOS) for the HP-I ($C2/m$, 10 GPa) FePSe$_3$ phase are
shown for selected displacements  $\Delta z$(P)  in Fig.\ \ref{fig6}.

\section{\sffamily \Large RESULTS AND DISCUSSION}

\subsection{\sffamily \Large Structural, Electronic and Phonon Properties}

The calculated crystallographic parameters and band gap values for FePSe$_3$ 
in the LP (0 GPa), HP-I (10 GPa) and HP-II (20 GPa) phases are reported in Table\ \ref{table1}. In the LP phase, the obtained results show rather good agreement with the available experimental lattice parameters \cite{WIEDENMANN1981} and band gap values,\cite{Du2016} however, the value of the $c$-axis parameter is overestimated.  

The phonon frequencies were calculated to demonstrate the stability of three structures. The dimensions of the primitive unit cell do not change significantly  across the LP($V$=191~\AA$^3$)-to-HP-I($V$=193~\AA$^3$) and HP-I($V$=178~\AA$^3$)-to-HP-II($V$=173~\AA$^3$) phase transitions, therefore the phonon instabilities are expected to occur at the Brillouin zone center \cite{Baroni2001}. 
The obtained phonon frequencies for the Raman and infrared active modes as well as silent modes are summarized in Table\ \ref{table2}. 
The absence of any imaginary frequencies
indicates that the optimized structures in all three phases of FePSe$_3$ correspond to a minimum on the surface of potential energy, making it possible to use the topological analysis (see below).
 
According to group-theoretical analysis for space group $R\bar{3}$ in the LP phase,
the 5E$_g$ and 5A$_g$ even modes are Raman-active, whereas the 5E$_u$  and 5A$_u$ odd modes are infrared-active (two of them (1E$_u$ and 1A$_u$) are acoustic modes with zero frequency at the  $\Gamma$-point). The agreement between the theoretical and experimentally observed\cite{Scagliotti1987,Bernasconi1988} phonon frequencies in the LP phase is good (Table\ \ref{table2}).
In the HP-I phase with the space group $C2/m$, the 7B$_g$ and 8A$_g$ even modes are Raman-active, whereas the 9B$_u$  and 6A$_u$ odd modes are infrared-active (three of them (2B$_u$ and 1A$_u$) are acoustic modes with zero frequency at the  $\Gamma$-point).
In the HP-II phase with the space group $P\bar{3}1m$, the 5E$_g$ and 3A$_{1g}$ even modes  are Raman-active, whereas the 5E$_u$ and 4A$_{2u}$ odd modes are infrared-active (two of them (1E$_u$ and 1A$_{2u}$) are acoustic modes with zero frequency at the  $\Gamma$-point). There are  also three silent modes (1A$_{1u}$ and 2A$_{2g}$) in HP-II phase. To the best of our knowledge, the experimental frequencies for the HP-I and HP-II phases are not available yet. 
 
Pressure dependence of the band gap $E_g$ in the LP, HP-I, and HP-II phases was evaluated from the band structure calculations (Fig.\ \ref{fig3}) performed for the optimized FePSe$_3$ crystal geometry (lattice parameters and atomic fractional coordinates) and is shown in the lower panel in Fig.\ \ref{fig2}. The band gap values were also estimated for respective space groups  beyond the existence ranges of the HP-I and HP-II phases (shown by open circles in Fig.\ \ref{fig2}).   

The transition from the low-pressure trigonal $R\bar{3}$ phase to 
the monoclinic HP-I $C2/m$ phase at 6 GPa is accompanied by a reduction of the volume and the number of layers in the conventional unit cell. Besides, in the HP-I phase the sliding of
the adjacent layers occurs in the $ab$-plane leading to the positioning of the P--P atom pairs in the neighbouring layers on top of each other along the $c$-axis (Fig.\ \ref{fig1}). A variation in the band gap during the phase transition at 6 GPa does not show any abrupt jumps within the accuracy of our calculations.

The collapse of the band gap occurs in FePSe$_3$  during the SMT from the HP-I to HP-II phase  at $\sim$13 GPa (Fig.\ \ref{fig2}). It is accompanied by a reduction of the unit cell volume by $\sim$3\% and of the vdW gap $d$ by $\sim$8\%. Note that these numbers are about twice smaller than for FePS$_3$.\cite{Evarestov2020}  
A simulated  compression of the FePSe$_3$ lattice with the space group $C2/m$ under pressure up to 30 GPa  as well as the simulated behaviour of the FePSe$_3$ lattice with the space group $P\bar{3}1m$ below 13 GPa do not result in any SMT (open circles in the lower panel in  Fig.\ \ref{fig2}). 
It is important that in the HP-II phase phosphorous atoms are displaced  
along the $c$-axis towards the vdW gap, so that they are located almost in the same plane together with selenium atoms. Similar behaviour under compression has been observed by us in FePS$_3$ where phosphorous atoms move to the sulfur plane.\cite{Evarestov2020} Such displacement leads to an increase of the distance 
between two intralayer phosphorous atoms P$_1$ and P$_2$ 
and a decrease of the distance between  two interlayer phosphorous atoms  P$_1$ and P$_3$ (Fig.\ \ref{fig1}), thus modifying the interaction strength between them. This situation will be addressed in detail below using the topological analysis of the electron density. 

The change of the electronic structure of FePSe$_3$ under pressure can be evidenced  following its band structure (Fig.\ \ref{fig3}) and the projected density of states (DOS) (Fig.\ \ref{fig4}).
The top of the valence bands in the LP and HP-I phases is mainly composed of the 3d(Fe) and 4p(Se) states, whereas the bottom of the conduction bands originates mainly from  3d(Fe), 3p(P) and 4p(Se) states.  The contribution of the phosphorus atoms becomes particularly important in the HP-II phase, similarly to the case of FePS$_3$.\cite{Evarestov2020} 

Again, we found that the position of the phosphorus atoms, relative to the layer composed of the selenium atoms, is crucial for SMT. In Fig.\ \ref{fig5}, we present three structural models constructed using the optimized crystallographic structures for the LP (0 GPa), HP-I (10 GPa), and HP-II (20 GPa) phases. In these models, 
phosphorus atoms were displaced along the $c$-axis direction by $\Delta z$(P) ($-0.1$ \AA\ $\leq \Delta z$(P)$ \leq 0.6$ \AA) relative to their optimized positions. We found that the sufficiently large displacement of the phosphorus atoms  in the direction of the plane formed by selenium atoms in the LP and HP-I phases (Fig.\ \ref{fig1}) leads to a decrease of the band gap and a transition to the metallic state. At the same time, the displacement of phosphorus atoms in the direction opposite to the  
plane formed by selenium atoms opens the gap in the HP-II phase for $\Delta z$(P)$ > 0.3$ \AA. 

The effect of the phosphorus atom displacements on the total and partial DOS is shown in Fig.\ \ref{fig6} for the case of the HP-I phase at 10 GPa. 
As one can see, the electronic structure of FePSe$_3$ is determined by  
hybridization of the 3d(Fe), 3p(P), and 3p(S) states around the Fermi level. 
An increase of pressure leads to the relative displacement of the phosphorous atoms   and promotes the broadening of both valence and conduction bands and a collapse of the band gap, i.e., insulator-to-metal transition. Thus, the SMT can be tuned by controlling the relative displacements of the phosphorous  atoms.

\subsection{\sffamily \Large Topological Analysis of P-P and Se-Se Chemical Bonds}

The nature of chemical bonding in molecules and solids is defined by the electron density $\rho(\textbf{r})$, conventionally considered to be weakly dependent on the calculation details such as, for example, the choice of the atomic basis and DFT functionals.
This assumption is confirmed in the present work by comparing the results of the topological analysis based on the B3LYP-13\%, PBE0 and M06 DFT calculations.

The topological analysis of the electron density of FePSe$_3$ was 
performed based on the Quantum Theory of Atoms in Molecules (QTAIM), due to
Richard Bader and coworkers,\cite{Bader1994} which is currently applied both to gas-phase molecules and solids.
As a result of the analysis, a set of critical points (CPs), at which the density gradient vanishes ($\bigtriangledown \rho(\textbf{r})=0$), was obtained and 
classified in terms of their type ($r$,$s$), where $r$ is the rank, and $s$ is the signature. Note that for stable three-dimensional structures, the critical points of $\rho(\textbf{r})$ have rank 3  and correspond to the elements of the chemically recognizable structures in a crystal such as nuclei (3,-3), bonds (3,-1), rings (3,+1) and cages (3,+3).\cite{Gatti1994} The crystal electron density in each of the three FePSe$_3$ phases has the symmetry of a corresponding space group. Therefore, the sites ensured to be a CP can be nothing but those Wyckoff positions (WP)\cite{Kroumova2003} of the corresponding space group.

While most of the QTAIM theory remains unaltered for solids, some significant differences should be taken into account.\cite{Luana2007}
The crystalline solid belongs to a three-dimensional torus.\cite{Luana2007} As a consequence,  the relationship, that connects the number
of critical points (CP) of any molecular scalar field ($n-b+r-c=1$), is replaced in the  crystals  by the  Morse relationship\cite{Morse1969}
$n-b+r-c=0$  for the crystal primitive unit cell. This relationship was used in the present work to check the total number of CPs in three phases of FePSe$_3$. 
    
The CPs of electron density found in the topological analysis for FePSe$_3$ is   related to WPs of the crystal space group, so that the WP multiplicity was used to determine the number of $n$, $b$, $r$ and $c$ type CPs.\cite{Gatti2005} 
The latter point is important since the search for CPs is realized by a procedure, depending on the search method and input parameters used in the TOPOND code.\cite{Gatti1994} 
                           
In Table\ \ref{table3} we report CPs and their WPs  for three phases of FePSe$_3$ crystal.
For example, in the space group  HP-II ($P\bar{3}1m$), Fe atoms (CP (3,-3)) occupy fixed WP 2c(1/3, 2/3, 0), whereas P and Se atoms (CPs (3,-3)) occupy 2e (one free-parameter) and 6k (two free-parameters) WPs with the local point symmetries C$_{\rm 3v}$ and C$_s$, respectively. One can see from Table\ \ref{table3} that two CPs (3,-1) and (3,+3) appear at the WP 6k. Evidently, all CPs at the  two-parameter  WP 6k(x, 0, z)  have different parameter values. 
Note that the obtained number of CPs fulfills the Morse relationship.\cite{Morse1969} 

The values of the electron density  $\rho(\textbf{r})$ and its Laplacian $\bigtriangledown^2 \rho$  at the CP (3,-1) were used to evaluate the type of the interatomic bonds in FePSe$_3$. 
In particular, the positive (negative) value of the Laplacian means that the electron density at the  point $\textbf{r}$ is lower (higher) in value than it is on average in an infinitesimal volume around  $\textbf{r}$. Therefore, regions with positive Laplacian are locally charge depleted, while regions with negative Laplacian have locally concentrated charge. 
The location of CP (3,-1) on the interatomic line is a necessary condition  for the existence of the chemical bond, whose classification into three groups (closed-shell (CS), shared-shell (SS) and intermediate-shell (IS)) was suggested in Refs.\  \cite{Cremer1984,Espinosa2002,Marabello2004,Tsirelson2014}.

Table\ \ref{table4} shows the results of  the topological analysis of the electron density and its Laplacian for selected P--P and Se--Se  atom pairs in three FePSe$_3$  phases, corresponding to the different pressures.  
$R_1$ and  $R_2$  are the distances from the  CP (3,-1) to the bound atoms. It is seen that  their sum $R_1$+$R_2$   is close to the value of the interatomic distance  $R$ for all (3,-1) CP’s. This means that these CP’s  lie practically on the line connecting two atoms.

In Table\ \ref{table4}, P$_1$(Se$_1$)--P$_2$(Se$_2$) and P$_1$(Se$_1$)--P$_3$(Se$_3$) are the intralayer and interlayer pairs of atoms, respectively.
It is seen that at pressures just above the SMT (13 GPa and 20 GPa)  the electron density at CP’s  of  interlayer P$_1$--P$_3$ bonds  is  larger than that for intralayer P$_1$--P$_2$ bonds and both Se--Se interlayer bonds.  
Also, the Laplacian of the electron density for the P$_1$--P$_3$ interlayer bonds  is negative,  i.e.,  the electron charge is locally concentrated at the bond, whereas the Laplacian is positive for the P$_1$--P$_2$ intralayer and both Se--Se interlayer bonds. However, a further increase in pressure to 30 GPa leads to equalization of the electron density between the P$_1$--P$_2$ and P$_1$--P$_3$ bonds and the 
recovery of the P$_1$--P$_2$ intralayer bonds, as is evidenced by the negative Laplacian value. Thus, the results of the topological analysis demonstrate that the SMT is related to an increase  of the electron density between 
P$_1$ and P$_3$ atoms located in the different neighbouring layers, leading to the formation 
of the P$_1$--P$_3$ chemical bonds.

Besides the Laplacian, several other bond descriptors were evaluated for selected P--P and Se--Se bond CPs of FePSe$_3$ (Table\ \ref{table5}). 
These are the potential energy density $V(\textbf{r})$, the positive definite kinetic energy density $G(\textbf{r})$ and the total electronic energy density
$H(\textbf{r}) = V(\textbf{r}) + G(\textbf{r})$.\cite{Gatti2005}
Using these descriptors, the nature of bonds can be described.\cite{Gatti2005}	
 Negative values of Laplacian and $H(\textbf{r})$ and $|V(\textbf{r})| / G(\textbf{r}) >2$ are attributed to the SS interactions such as covalent or polar bonds.
Intermediate bonds are associated with IS interactions, which are described by  positive Laplacian, an almost zero value of the total electronic energy density $H(\textbf{r})$  and $1 < |V(\textbf{r})| / G(\textbf{r}) < 2$. 
Positive Laplacian and $H(\textbf{r})$ and $|V(\textbf{r})| / G(\textbf{r}) < 1$  are indicative for the CS interactions such as ionic, hydrogen bonds and van der Waals interactions. 

Detailed analysis of CPs for P$_1$(Se$_1$)--P$_2$(Se$_2$) and P$_1$(Se$_1$)--P$_3$(Se$_3$) bonds at three pressures $P$ = 0, 10, and 20 GPa is given in Table\ \ref{table5}.
One can classify  P$_1$(Se$_1$)--P$_2$(Se$_2$) and   P$_1$(Se$_1$)--P$_3$(Se$_3$) bonds over  CS, SS and IS types at three selected pressures. According to the classification scheme, described above, the intralayer P$_1$--P$_2$ bonds can be described as the shared-shell covalent bonds in the LP and HP-I phases (at the pressures 0 GPa and 10 GPa, respectively) with negative  $\bigtriangledown^2 \rho(\textbf{r})$, $|V(\textbf{r})| / G(\textbf{r}) > 2$, negative total energy density 
$H(\rho(\textbf{r}))$  and the kinetic energy per electron $G(\textbf{r})/\rho(\textbf{r}) < 1$. In the HP-II phase (at 20 GPa), the P$_1$--P$_2$ bond type change to the intermediate shell bond with positive  $\bigtriangledown^2 \rho(\textbf{r})$ and $1< |V(\textbf{r})| / G(\textbf{r}) <2$. The total energy density $H(\rho(\textbf{r}))$ remains negative (but becomes about five times smaller), while  $G(\textbf{r})/\rho(\textbf{r})$  remains mostly the same. Note that the interlayer P$_1$--P$_3$ bonds are absent in the LP and HP-I phases but appear in the HP-II phase and can be classified as the shared-shell covalent bonds ($\bigtriangledown^2 \rho(\textbf{r}) < 0$, $|V(\textbf{r})| / G(\textbf{r}) >2$, $H(\rho(\textbf{r})) < 0$ and $ G(\textbf{r})/\rho(\textbf{r}) <1$), which are weaker than  the P$_1$--P$_2$ SS bonds for two  lower pressures.
The interlayer Se$_1$--Se$_2$  and  Se$_1$--Se$_3$ bonds also change  with increasing pressure and can be classified as the intermediate shell bonds:   $\bigtriangledown^2 \rho(\textbf{r})$ is always small, but remains positive;  $|V(\textbf{r})| / G(\textbf{r})$  grows from 0.7  to 1.1 with increasing pressure, the total energy density $H(\rho(\textbf{r}))$ is very small positive in the LP and HP-I phases at the two smaller pressures and becomes negative in the HP-II phase at 20 GPa; $ G(\textbf{r})/\rho(\textbf{r})$ is less than  1 but larger than for the P--P bonds.

The results of the topological analysis for the PBE0 and M06 functionals  are shown in Table\ \ref{table6}. They are qualitatively similar to those obtained using the B3LYP-13\% functional thus allowing to withdraw the same conclusions regarding the chemical bonding in FePSe$_3$.

The electron localization function (ELF) contour maps for LP (0 GPa), HP-I (10 GPa), and HP-II (20 GPa) phases of FePSe$_3$ are shown in Fig.\ \ref{fig7} for selected planes, being orthogonal to the vdW gaps and containing P--P atom pairs.
The vdW gaps with low electron density are well observed in the LP phase between layers. 
The P--P atom pairs located in the neighbouring layers are displaced along the $b$-axis that makes  P$_1$ and P$_3$ phosphorus atoms well separated (Fig.\ \ref{fig1}). 
Upon phase transition to the HP-I phase, the adjacent layers displace relative to each other and come closer, however, they are still separated by the vdW gaps, visible in Fig.\ \ref{fig7}. While the P--P atom pairs located in the neighbouring layers are situated on top of each other along the $c$-axis, they are now displaced along the $a$-axis (therefore P$_3$ atom is not visible in the HP-I phase in Fig.\ \ref{fig7}), thus preventing the bonding between the  P$_1$ and P$_3$ phosphorus atoms  (see also Fig.\ \ref{fig1}). 
Finally, upon SMT at 13 GPa, the van der Waals gaps vanish, and P$_1$ and P$_3$ phosphorus atoms, located in LP and HP-I phases in different layers, become strongly bound along the $c$-axis. This fact is well observed in the ELF contour map of the HP-II phase as an increase of the electron 
localization in the interatomic region between the two phosphorus P$_1$ and P$_3$ located in different layers, which is accompanied by a slight reduction of electron localization between P$_1$ and P$_2$ atoms located in the same layer.

\section{\sffamily \Large CONCLUSIONS}

The behaviour of the structural, electronic and phonon properties in the layered iron phosphorus triselenide (FePSe$_3$) under pressure has been theoretically studied 
based on the first-principles LCAO calculations using hybrid DFT-HF B3LYP functional. 
Detailed analysis of the chemical bonding in FePSe$_3$ has been performed using the topological analysis of the electron density, according to the Quantum Theory of Atoms in Molecules.\cite{Bader1994}

Two pressure-induced phase transitions have been evidenced at about 6 GPa and 13 GPa, 
in agreement with the recent experimental\cite{Wang2018} and theoretical\cite{Zheng2019} findings. At ambient pressure,  FePSe$_3$ has 
the indirect band gap of about 1.6 eV, compared to the experimental value of  
1.3 eV.\cite{Brec1979}
The first phase transition  is accompanied by a change in symmetry  from $R\bar{3}$ to $C2/m$,
 sliding of adjacent layers in the $ab$-plane and a small reduction of the band gap 
down to about 1.2-1.4 eV. At the second phase transition from $C2/m$ to $P\bar{3}1m$
phase, the band gap collapses, and FePSe$_3$ transforms to the metallic state. 
Moreover, similar to the case of FePS$_3$,\cite{Evarestov2020} metallic conductivity 
in FePSe$_3$ could also occur in the LP and HP-I phases if the P and Se atoms were located 
almost in the same plane.

The analysis of the projected density of states suggests that the 3p states of phosphorus atoms contribute significantly to the bottom of the conduction band.  
Therefore, the collapse of the band gap upon the SMT at about 13 GPa is caused by 
changes in the electronic structure of FePSe$_3$ induced by relative displacements of phosphorus or selenium atoms along the $c$-axis direction under pressure.
The topological analysis of the electron density and its  Laplacian in FePSe$_3$  
suggests that the pressure affects not only interatomic distances but the bond nature  between the intralayer and interlayer phosphorus atoms. In particular, the type of the intralayer P--P bonds changes from the shared-shell covalent bonds in the LP and HP-I phases to the intermediate-shell bond in the HP-II phase. At the same time, the interlayer P--P bonds, absent in the LP and HP-I phases, appear in the HP-II phase as the shared-shell covalent bonds.

\subsection*{\sffamily \large ACKNOWLEDGMENTS}
The authors acknowledge the assistance of the University Computer
Center of Saint-Petersburg State University in the accomplishment
of high-performance computations.
A.K. is grateful to the Latvian Council of Science project no. lzp-2018/2-0353 for financial support.
Institute of Solid State Physics, University of Latvia as the Center of Excellence has received funding from the European Union's Horizon 2020 Framework Programme H2020-WIDESPREAD-01-2016-2017-TeamingPhase2 under grant agreement No. 739508, project CAMART2.


\clearpage



\begin{thebibliography}{67}
	\expandafter\ifx\csname natexlab\endcsname\relax\def\natexlab#1{#1}\fi
	\expandafter\ifx\csname bibnamefont\endcsname\relax
	\def\bibnamefont#1{#1}\fi
	\expandafter\ifx\csname bibfnamefont\endcsname\relax
	\def\bibfnamefont#1{#1}\fi
	\expandafter\ifx\csname citenamefont\endcsname\relax
	\def\citenamefont#1{#1}\fi
	\expandafter\ifx\csname url\endcsname\relax
	\def\url#1{\texttt{#1}}\fi
	\expandafter\ifx\csname urlprefix\endcsname\relax\def\urlprefix{URL }\fi
	\providecommand{\bibinfo}[2]{#2}
	\providecommand{\eprint}[2][]{\url{#2}}
	
	\bibitem[{\citenamefont{Bader}(1994)}]{Bader1994}
	\bibinfo{author}{\bibfnamefont{R.~F.~W.} \bibnamefont{Bader}},
	\emph{\bibinfo{title}{Atoms in Molecules: A Quantum Theory}}
	(\bibinfo{publisher}{Clarendon Press}, \bibinfo{address}{Oxford},
	\bibinfo{year}{1994}).
	
	\bibitem[{\citenamefont{Lepetit et~al.}(2017)\citenamefont{Lepetit, Fau,
			Fajerwerg, Kahn, and Silvi}}]{LEPETIT2017}
	\bibinfo{author}{\bibfnamefont{C.}~\bibnamefont{Lepetit}},
	\bibinfo{author}{\bibfnamefont{P.}~\bibnamefont{Fau}},
	\bibinfo{author}{\bibfnamefont{K.}~\bibnamefont{Fajerwerg}},
	\bibinfo{author}{\bibfnamefont{M.~L.} \bibnamefont{Kahn}}, \bibnamefont{and}
	\bibinfo{author}{\bibfnamefont{B.}~\bibnamefont{Silvi}},
	\bibinfo{journal}{Coord. Chem. Rev.} \textbf{\bibinfo{volume}{345}},
	\bibinfo{pages}{150} (\bibinfo{year}{2017}).
	
	\bibitem[{\citenamefont{Vyboishchikov et~al.}(1997)\citenamefont{Vyboishchikov,
			Sierraalta, and Frenking}}]{Vyboishchikov1997}
	\bibinfo{author}{\bibfnamefont{S.~F.} \bibnamefont{Vyboishchikov}},
	\bibinfo{author}{\bibfnamefont{A.}~\bibnamefont{Sierraalta}},
	\bibnamefont{and} \bibinfo{author}{\bibfnamefont{G.}~\bibnamefont{Frenking}},
	\bibinfo{journal}{J. Comput. Chem.} \textbf{\bibinfo{volume}{18}},
	\bibinfo{pages}{416} (\bibinfo{year}{1997}).
	
	\bibitem[{\citenamefont{Gatti}(2005)}]{Gatti2005}
	\bibinfo{author}{\bibfnamefont{C.}~\bibnamefont{Gatti}}, \bibinfo{journal}{Z.
		Kristallogr.} \textbf{\bibinfo{volume}{220}}, \bibinfo{pages}{399}
	(\bibinfo{year}{2005}).
	
	\bibitem[{\citenamefont{Johnson et~al.}(2010)\citenamefont{Johnson, Keinan,
			Mori-Sánchez, Contreras-García, Cohen, and Yang}}]{Johnson2010}
	\bibinfo{author}{\bibfnamefont{E.~R.} \bibnamefont{Johnson}},
	\bibinfo{author}{\bibfnamefont{S.}~\bibnamefont{Keinan}},
	\bibinfo{author}{\bibfnamefont{P.}~\bibnamefont{Mori-Sánchez}},
	\bibinfo{author}{\bibfnamefont{J.}~\bibnamefont{Contreras-García}},
	\bibinfo{author}{\bibfnamefont{A.~J.} \bibnamefont{Cohen}}, \bibnamefont{and}
	\bibinfo{author}{\bibfnamefont{W.}~\bibnamefont{Yang}}, \bibinfo{journal}{J.
		Am. Chem. Soc.} \textbf{\bibinfo{volume}{132}}, \bibinfo{pages}{6498}
	(\bibinfo{year}{2010}).
	
	\bibitem[{\citenamefont{Matta and Arabi}(2011)}]{Matta2011}
	\bibinfo{author}{\bibfnamefont{C.~F.} \bibnamefont{Matta}} \bibnamefont{and}
	\bibinfo{author}{\bibfnamefont{A.~A.} \bibnamefont{Arabi}},
	\bibinfo{journal}{Future Med. Chem.} \textbf{\bibinfo{volume}{3}},
	\bibinfo{pages}{969} (\bibinfo{year}{2011}).
	
	\bibitem[{\citenamefont{Krawczuk and Macchi}(2014)}]{Krawczuk2014}
	\bibinfo{author}{\bibfnamefont{A.}~\bibnamefont{Krawczuk}} \bibnamefont{and}
	\bibinfo{author}{\bibfnamefont{P.}~\bibnamefont{Macchi}},
	\bibinfo{journal}{Chem. Cent. J.} \textbf{\bibinfo{volume}{8}},
	\bibinfo{pages}{68} (\bibinfo{year}{2014}).
	
	\bibitem[{\citenamefont{Marana et~al.}(2016)\citenamefont{Marana, Casassa,
			Longo, and Sambrano}}]{Marana2016}
	\bibinfo{author}{\bibfnamefont{N.~L.} \bibnamefont{Marana}},
	\bibinfo{author}{\bibfnamefont{S.}~\bibnamefont{Casassa}},
	\bibinfo{author}{\bibfnamefont{E.}~\bibnamefont{Longo}}, \bibnamefont{and}
	\bibinfo{author}{\bibfnamefont{J.~R.} \bibnamefont{Sambrano}},
	\bibinfo{journal}{J. Phys. Chem. C} \textbf{\bibinfo{volume}{120}},
	\bibinfo{pages}{6814} (\bibinfo{year}{2016}).
	
	\bibitem[{\citenamefont{Dittrich}(2017)}]{Dittrich2017}
	\bibinfo{author}{\bibfnamefont{B.}~\bibnamefont{Dittrich}},
	\bibinfo{journal}{Acta Crystallogr. B} \textbf{\bibinfo{volume}{73}},
	\bibinfo{pages}{325} (\bibinfo{year}{2017}).
	
	\bibitem[{\citenamefont{Macchi}(2017)}]{Macchi2017}
	\bibinfo{author}{\bibfnamefont{P.}~\bibnamefont{Macchi}},
	\bibinfo{journal}{Acta Crystallogr. B} \textbf{\bibinfo{volume}{73}},
	\bibinfo{pages}{330} (\bibinfo{year}{2017}).
	
	\bibitem[{\citenamefont{Varadwaj et~al.}(2019)\citenamefont{Varadwaj, Varadwaj,
			Marques, and Yamashita}}]{Varadwaj2019}
	\bibinfo{author}{\bibfnamefont{P.~R.} \bibnamefont{Varadwaj}},
	\bibinfo{author}{\bibfnamefont{A.}~\bibnamefont{Varadwaj}},
	\bibinfo{author}{\bibfnamefont{H.~M.} \bibnamefont{Marques}},
	\bibnamefont{and}
	\bibinfo{author}{\bibfnamefont{K.}~\bibnamefont{Yamashita}},
	\bibinfo{journal}{Sci. Rep.} \textbf{\bibinfo{volume}{9}},
	\bibinfo{pages}{50} (\bibinfo{year}{2019}).
	
	\bibitem[{\citenamefont{Titah et~al.}(2019)\citenamefont{Titah, Ngwa, Sirsch,
			Karime, and Kone}}]{Titah2019}
	\bibinfo{author}{\bibfnamefont{J.~T.} \bibnamefont{Titah}},
	\bibinfo{author}{\bibfnamefont{F.~C.} \bibnamefont{Ngwa}},
	\bibinfo{author}{\bibfnamefont{P.}~\bibnamefont{Sirsch}},
	\bibinfo{author}{\bibfnamefont{C.~W.} \bibnamefont{Karime}},
	\bibnamefont{and} \bibinfo{author}{\bibfnamefont{M.~G.~R.}
		\bibnamefont{Kone}}, \bibinfo{journal}{Int. J. Comput. Theor. Chem.}
	\textbf{\bibinfo{volume}{7}}, \bibinfo{pages}{115} (\bibinfo{year}{2019}).
	
	\bibitem[{\citenamefont{Lacerda et~al.}(2019)\citenamefont{Lacerda, Ribeiro,
			and {de Lazaro}}}]{LACERDA2019}
	\bibinfo{author}{\bibfnamefont{L.~H.~S.} \bibnamefont{Lacerda}},
	\bibinfo{author}{\bibfnamefont{R.~A.~P.} \bibnamefont{Ribeiro}},
	\bibnamefont{and} \bibinfo{author}{\bibfnamefont{S.~R.} \bibnamefont{{de
				Lazaro}}}, \bibinfo{journal}{J. Magn. Magn. Mater.}
	\textbf{\bibinfo{volume}{480}}, \bibinfo{pages}{199} (\bibinfo{year}{2019}).
	
	\bibitem[{\citenamefont{Korabel{'}nikov and
			Zhuravlev}(2019)}]{Korabelnikov2019}
	\bibinfo{author}{\bibfnamefont{D.}~\bibnamefont{Korabel{'}nikov}}
	\bibnamefont{and} \bibinfo{author}{\bibfnamefont{Y.~N.}
		\bibnamefont{Zhuravlev}}, \bibinfo{journal}{RSC Adv.}
	\textbf{\bibinfo{volume}{9}}, \bibinfo{pages}{12020} (\bibinfo{year}{2019}).
	
	\bibitem[{\citenamefont{Tolborg and Iversen}(2019)}]{Tolborg2019}
	\bibinfo{author}{\bibfnamefont{K.}~\bibnamefont{Tolborg}} \bibnamefont{and}
	\bibinfo{author}{\bibfnamefont{B.~B.} \bibnamefont{Iversen}},
	\bibinfo{journal}{Chem. Eur. J.} \textbf{\bibinfo{volume}{25}},
	\bibinfo{pages}{15010} (\bibinfo{year}{2019}).
	
	\bibitem[{\citenamefont{Parisi et~al.}(2012)\citenamefont{Parisi, Sciascia,
			Princivalle, and Merli}}]{Parisi2012}
	\bibinfo{author}{\bibfnamefont{F.}~\bibnamefont{Parisi}},
	\bibinfo{author}{\bibfnamefont{L.}~\bibnamefont{Sciascia}},
	\bibinfo{author}{\bibfnamefont{F.}~\bibnamefont{Princivalle}},
	\bibnamefont{and} \bibinfo{author}{\bibfnamefont{M.}~\bibnamefont{Merli}},
	\bibinfo{journal}{Phys. Chem. Minerals} \textbf{\bibinfo{volume}{39}},
	\bibinfo{pages}{103} (\bibinfo{year}{2012}).
	
	\bibitem[{\citenamefont{Merli and Sciascia}(2013)}]{Merli2013}
	\bibinfo{author}{\bibfnamefont{M.}~\bibnamefont{Merli}} \bibnamefont{and}
	\bibinfo{author}{\bibfnamefont{L.}~\bibnamefont{Sciascia}},
	\bibinfo{journal}{Phys. Chem. Minerals} \textbf{\bibinfo{volume}{40}},
	\bibinfo{pages}{455} (\bibinfo{year}{2013}).
	
	\bibitem[{\citenamefont{Salvad\'{o} et~al.}(2013)\citenamefont{Salvad\'{o},
			Pertierra, Morales-Garc\'{i}a, Menéndez, and Recio}}]{Salvado2013}
	\bibinfo{author}{\bibfnamefont{M.~A.} \bibnamefont{Salvad\'{o}}},
	\bibinfo{author}{\bibfnamefont{P.}~\bibnamefont{Pertierra}},
	\bibinfo{author}{\bibfnamefont{A.}~\bibnamefont{Morales-Garc\'{i}a}},
	\bibinfo{author}{\bibfnamefont{J.~M.} \bibnamefont{Menéndez}},
	\bibnamefont{and} \bibinfo{author}{\bibfnamefont{J.~M.} \bibnamefont{Recio}},
	\bibinfo{journal}{J. Phys. Chem. C} \textbf{\bibinfo{volume}{117}},
	\bibinfo{pages}{8950} (\bibinfo{year}{2013}).
	
	\bibitem[{\citenamefont{Casati et~al.}(2016)\citenamefont{Casati, Kleppe,
			Jephcoat, and Macchi}}]{Casati2016}
	\bibinfo{author}{\bibfnamefont{N.}~\bibnamefont{Casati}},
	\bibinfo{author}{\bibfnamefont{A.}~\bibnamefont{Kleppe}},
	\bibinfo{author}{\bibfnamefont{A.~P.} \bibnamefont{Jephcoat}},
	\bibnamefont{and} \bibinfo{author}{\bibfnamefont{P.}~\bibnamefont{Macchi}},
	\bibinfo{journal}{Nat. Commun.} \textbf{\bibinfo{volume}{7}},
	\bibinfo{pages}{10901} (\bibinfo{year}{2016}).
	
	\bibitem[{\citenamefont{Pereira et~al.}(2016)\citenamefont{Pereira, Gomis,
			Sans, Contreras-Garc\'{\i}a, Manj\'on, Rodr\'{\i}guez-Hern\'andez, Mu\~noz,
			and Beltr\'an}}]{Pereira2016}
	\bibinfo{author}{\bibfnamefont{A.~L.~J.} \bibnamefont{Pereira}},
	\bibinfo{author}{\bibfnamefont{O.}~\bibnamefont{Gomis}},
	\bibinfo{author}{\bibfnamefont{J.~A.} \bibnamefont{Sans}},
	\bibinfo{author}{\bibfnamefont{J.}~\bibnamefont{Contreras-Garc\'{\i}a}},
	\bibinfo{author}{\bibfnamefont{F.~J.} \bibnamefont{Manj\'on}},
	\bibinfo{author}{\bibfnamefont{P.}~\bibnamefont{Rodr\'{\i}guez-Hern\'andez}},
	\bibinfo{author}{\bibfnamefont{A.}~\bibnamefont{Mu\~noz}}, \bibnamefont{and}
	\bibinfo{author}{\bibfnamefont{A.}~\bibnamefont{Beltr\'an}},
	\bibinfo{journal}{Phys. Rev. B} \textbf{\bibinfo{volume}{93}},
	\bibinfo{pages}{224111} (\bibinfo{year}{2016}).
	
	\bibitem[{\citenamefont{Merli and Pavese}(2018)}]{Merli2018}
	\bibinfo{author}{\bibfnamefont{M.}~\bibnamefont{Merli}} \bibnamefont{and}
	\bibinfo{author}{\bibfnamefont{A.}~\bibnamefont{Pavese}},
	\bibinfo{journal}{Acta Crystallogr. A} \textbf{\bibinfo{volume}{74}},
	\bibinfo{pages}{102} (\bibinfo{year}{2018}).
	
	\bibitem[{\citenamefont{Belarouci et~al.}(2018)\citenamefont{Belarouci,
			Ouahrani, Benabdallah, Morales-Garc\'{i}a, and Franco}}]{Belarouci2018}
	\bibinfo{author}{\bibfnamefont{S.}~\bibnamefont{Belarouci}},
	\bibinfo{author}{\bibfnamefont{T.}~\bibnamefont{Ouahrani}},
	\bibinfo{author}{\bibfnamefont{N.}~\bibnamefont{Benabdallah}},
	\bibinfo{author}{\bibfnamefont{A.}~\bibnamefont{Morales-Garc\'{i}a}},
	\bibnamefont{and} \bibinfo{author}{\bibfnamefont{R.}~\bibnamefont{Franco}},
	\bibinfo{journal}{Phase Transit.} \textbf{\bibinfo{volume}{91}},
	\bibinfo{pages}{759} (\bibinfo{year}{2018}).
	
	\bibitem[{\citenamefont{Schwarz et~al.}(2019)\citenamefont{Schwarz, Wosylus,
			Schmidt, Akselrud, Ormeci, Hanfland, Hermann, and Kuntscher}}]{Schwarz2019}
	\bibinfo{author}{\bibfnamefont{U.}~\bibnamefont{Schwarz}},
	\bibinfo{author}{\bibfnamefont{A.}~\bibnamefont{Wosylus}},
	\bibinfo{author}{\bibfnamefont{M.}~\bibnamefont{Schmidt}},
	\bibinfo{author}{\bibfnamefont{L.}~\bibnamefont{Akselrud}},
	\bibinfo{author}{\bibfnamefont{A.}~\bibnamefont{Ormeci}},
	\bibinfo{author}{\bibfnamefont{M.}~\bibnamefont{Hanfland}},
	\bibinfo{author}{\bibfnamefont{V.}~\bibnamefont{Hermann}}, \bibnamefont{and}
	\bibinfo{author}{\bibfnamefont{C.}~\bibnamefont{Kuntscher}},
	\bibinfo{journal}{Inorganics} \textbf{\bibinfo{volume}{7}},
	\bibinfo{pages}{143} (\bibinfo{year}{2019}).
	
	\bibitem[{\citenamefont{Meyer et~al.}(2019)\citenamefont{Meyer, Barthel, Mace,
			Vannay, Guillot, Smit, and Corminboeuf}}]{Meyer2019}
	\bibinfo{author}{\bibfnamefont{B.}~\bibnamefont{Meyer}},
	\bibinfo{author}{\bibfnamefont{S.}~\bibnamefont{Barthel}},
	\bibinfo{author}{\bibfnamefont{A.}~\bibnamefont{Mace}},
	\bibinfo{author}{\bibfnamefont{L.}~\bibnamefont{Vannay}},
	\bibinfo{author}{\bibfnamefont{B.}~\bibnamefont{Guillot}},
	\bibinfo{author}{\bibfnamefont{B.}~\bibnamefont{Smit}}, \bibnamefont{and}
	\bibinfo{author}{\bibfnamefont{C.}~\bibnamefont{Corminboeuf}},
	\bibinfo{journal}{J. Phys. Chem. Lett.} \textbf{\bibinfo{volume}{10}},
	\bibinfo{pages}{1482} (\bibinfo{year}{2019}).
	
	\bibitem[{\citenamefont{Parisi et~al.}(2019)\citenamefont{Parisi, Sciascia,
			Princivalle, and Merli}}]{Parisi2019}
	\bibinfo{author}{\bibfnamefont{F.}~\bibnamefont{Parisi}},
	\bibinfo{author}{\bibfnamefont{L.}~\bibnamefont{Sciascia}},
	\bibinfo{author}{\bibfnamefont{F.}~\bibnamefont{Princivalle}},
	\bibnamefont{and} \bibinfo{author}{\bibfnamefont{M.}~\bibnamefont{Merli}},
	\bibinfo{journal}{Ceram. Int.} \textbf{\bibinfo{volume}{45}},
	\bibinfo{pages}{2820} (\bibinfo{year}{2019}).
	
	\bibitem[{\citenamefont{Gajda et~al.}(2020)\citenamefont{Gajda, Stachowicz,
			Makal, Sutu{\l}a, Parafiniuk, Fertey, and Wo{\'{z}}niak}}]{Gajda2020}
	\bibinfo{author}{\bibfnamefont{R.}~\bibnamefont{Gajda}},
	\bibinfo{author}{\bibfnamefont{M.}~\bibnamefont{Stachowicz}},
	\bibinfo{author}{\bibfnamefont{A.}~\bibnamefont{Makal}},
	\bibinfo{author}{\bibfnamefont{S.}~\bibnamefont{Sutu{\l}a}},
	\bibinfo{author}{\bibfnamefont{J.}~\bibnamefont{Parafiniuk}},
	\bibinfo{author}{\bibfnamefont{P.}~\bibnamefont{Fertey}}, \bibnamefont{and}
	\bibinfo{author}{\bibfnamefont{K.}~\bibnamefont{Wo{\'{z}}niak}},
	\bibinfo{journal}{IUCrJ} \textbf{\bibinfo{volume}{7}}, \bibinfo{pages}{383}
	(\bibinfo{year}{2020}).
	
	\bibitem[{\citenamefont{Kasai et~al.}(2018)\citenamefont{Kasai, Tolborg, Sist,
			Zhang, Hathwar, Fils\o{}, Cenedese, Sugimoto, Overgaard, Nishibori
			et~al.}}]{Kasai2018}
	\bibinfo{author}{\bibfnamefont{H.}~\bibnamefont{Kasai}},
	\bibinfo{author}{\bibfnamefont{K.}~\bibnamefont{Tolborg}},
	\bibinfo{author}{\bibfnamefont{M.}~\bibnamefont{Sist}},
	\bibinfo{author}{\bibfnamefont{J.}~\bibnamefont{Zhang}},
	\bibinfo{author}{\bibfnamefont{V.~R.} \bibnamefont{Hathwar}},
	\bibinfo{author}{\bibfnamefont{M.~O.} \bibnamefont{Fils\o{}}},
	\bibinfo{author}{\bibfnamefont{S.}~\bibnamefont{Cenedese}},
	\bibinfo{author}{\bibfnamefont{K.}~\bibnamefont{Sugimoto}},
	\bibinfo{author}{\bibfnamefont{J.}~\bibnamefont{Overgaard}},
	\bibinfo{author}{\bibfnamefont{E.}~\bibnamefont{Nishibori}},
	\bibnamefont{et~al.}, \bibinfo{journal}{Nat. Mater.}
	\textbf{\bibinfo{volume}{17}}, \bibinfo{pages}{249} (\bibinfo{year}{2018}).
	
	\bibitem[{\citenamefont{Ryzhikov et~al.}(2014)\citenamefont{Ryzhikov, Slepkov,
			Kozlova, and Gabuda}}]{Ryzhikov2014}
	\bibinfo{author}{\bibfnamefont{M.~R.} \bibnamefont{Ryzhikov}},
	\bibinfo{author}{\bibfnamefont{V.~A.} \bibnamefont{Slepkov}},
	\bibinfo{author}{\bibfnamefont{S.~G.} \bibnamefont{Kozlova}},
	\bibnamefont{and} \bibinfo{author}{\bibfnamefont{S.~P.}
		\bibnamefont{Gabuda}}, \bibinfo{journal}{J. Comput. Chem.}
	\textbf{\bibinfo{volume}{35}}, \bibinfo{pages}{1641} (\bibinfo{year}{2014}).
	
	\bibitem[{\citenamefont{Wang et~al.}(2018)\citenamefont{Wang, Ying, Zhou, Sun,
			Wen, Zhou, Li, Zhang, Han, Xiao et~al.}}]{Wang2018}
	\bibinfo{author}{\bibfnamefont{Y.}~\bibnamefont{Wang}},
	\bibinfo{author}{\bibfnamefont{J.}~\bibnamefont{Ying}},
	\bibinfo{author}{\bibfnamefont{Z.}~\bibnamefont{Zhou}},
	\bibinfo{author}{\bibfnamefont{J.}~\bibnamefont{Sun}},
	\bibinfo{author}{\bibfnamefont{T.}~\bibnamefont{Wen}},
	\bibinfo{author}{\bibfnamefont{Y.}~\bibnamefont{Zhou}},
	\bibinfo{author}{\bibfnamefont{N.}~\bibnamefont{Li}},
	\bibinfo{author}{\bibfnamefont{Q.}~\bibnamefont{Zhang}},
	\bibinfo{author}{\bibfnamefont{F.}~\bibnamefont{Han}},
	\bibinfo{author}{\bibfnamefont{Y.}~\bibnamefont{Xiao}}, \bibnamefont{et~al.},
	\bibinfo{journal}{Nat. Commun.} \textbf{\bibinfo{volume}{9}},
	\bibinfo{pages}{1914} (\bibinfo{year}{2018}).
	
	\bibitem[{\citenamefont{Brec et~al.}(1979)\citenamefont{Brec, Schleich,
			Ouvrard, Louisy, and Rouxel}}]{Brec1979}
	\bibinfo{author}{\bibfnamefont{R.}~\bibnamefont{Brec}},
	\bibinfo{author}{\bibfnamefont{D.~M.} \bibnamefont{Schleich}},
	\bibinfo{author}{\bibfnamefont{G.}~\bibnamefont{Ouvrard}},
	\bibinfo{author}{\bibfnamefont{A.}~\bibnamefont{Louisy}}, \bibnamefont{and}
	\bibinfo{author}{\bibfnamefont{J.}~\bibnamefont{Rouxel}},
	\bibinfo{journal}{Inorg. Chem.} \textbf{\bibinfo{volume}{18}},
	\bibinfo{pages}{1814} (\bibinfo{year}{1979}).
	
	\bibitem[{\citenamefont{Taylor et~al.}(1974)\citenamefont{Taylor, Steger, Wold,
			and Kostiner}}]{Taylor1974}
	\bibinfo{author}{\bibfnamefont{B.}~\bibnamefont{Taylor}},
	\bibinfo{author}{\bibfnamefont{J.}~\bibnamefont{Steger}},
	\bibinfo{author}{\bibfnamefont{A.}~\bibnamefont{Wold}}, \bibnamefont{and}
	\bibinfo{author}{\bibfnamefont{E.}~\bibnamefont{Kostiner}},
	\bibinfo{journal}{Inorg. Chem.} \textbf{\bibinfo{volume}{13}},
	\bibinfo{pages}{2719} (\bibinfo{year}{1974}).
	
	\bibitem[{\citenamefont{Wiedenmann et~al.}(1981)\citenamefont{Wiedenmann,
			Rossat-Mignod, Louisy, Brec, and Rouxel}}]{WIEDENMANN1981}
	\bibinfo{author}{\bibfnamefont{A.}~\bibnamefont{Wiedenmann}},
	\bibinfo{author}{\bibfnamefont{J.}~\bibnamefont{Rossat-Mignod}},
	\bibinfo{author}{\bibfnamefont{A.}~\bibnamefont{Louisy}},
	\bibinfo{author}{\bibfnamefont{R.}~\bibnamefont{Brec}}, \bibnamefont{and}
	\bibinfo{author}{\bibfnamefont{J.}~\bibnamefont{Rouxel}},
	\bibinfo{journal}{Solid State Commun.} \textbf{\bibinfo{volume}{40}},
	\bibinfo{pages}{1067} (\bibinfo{year}{1981}).
	
	\bibitem[{\citenamefont{{Le Flem} et~al.}(1982)\citenamefont{{Le Flem}, Brec,
			Ouvard, Louisy, and Segransan}}]{LeFlem1982}
	\bibinfo{author}{\bibfnamefont{G.}~\bibnamefont{{Le Flem}}},
	\bibinfo{author}{\bibfnamefont{R.}~\bibnamefont{Brec}},
	\bibinfo{author}{\bibfnamefont{G.}~\bibnamefont{Ouvard}},
	\bibinfo{author}{\bibfnamefont{A.}~\bibnamefont{Louisy}}, \bibnamefont{and}
	\bibinfo{author}{\bibfnamefont{P.}~\bibnamefont{Segransan}},
	\bibinfo{journal}{J. Phys. Chem. Solids} \textbf{\bibinfo{volume}{43}},
	\bibinfo{pages}{455} (\bibinfo{year}{1982}).
	
	\bibitem[{\citenamefont{Zheng et~al.}(2019)\citenamefont{Zheng, Jiang, Xue,
			Dai, and Feng}}]{Zheng2019}
	\bibinfo{author}{\bibfnamefont{Y.}~\bibnamefont{Zheng}},
	\bibinfo{author}{\bibfnamefont{X.-X.} \bibnamefont{Jiang}},
	\bibinfo{author}{\bibfnamefont{X.-X.} \bibnamefont{Xue}},
	\bibinfo{author}{\bibfnamefont{J.}~\bibnamefont{Dai}}, \bibnamefont{and}
	\bibinfo{author}{\bibfnamefont{Y.}~\bibnamefont{Feng}},
	\bibinfo{journal}{Phys. Rev. B} \textbf{\bibinfo{volume}{100}},
	\bibinfo{pages}{174102} (\bibinfo{year}{2019}).
	
	\bibitem[{\citenamefont{Gusm\~{a}o et~al.}(2017)\citenamefont{Gusm\~{a}o,
			Sofer, Sedmidubsk\'{y}, Huber, and Pumera}}]{Gusmao2017}
	\bibinfo{author}{\bibfnamefont{R.}~\bibnamefont{Gusm\~{a}o}},
	\bibinfo{author}{\bibfnamefont{Z.}~\bibnamefont{Sofer}},
	\bibinfo{author}{\bibfnamefont{D.}~\bibnamefont{Sedmidubsk\'{y}}},
	\bibinfo{author}{\bibfnamefont{{\v{S}}.}~\bibnamefont{Huber}},
	\bibnamefont{and} \bibinfo{author}{\bibfnamefont{M.}~\bibnamefont{Pumera}},
	\bibinfo{journal}{ACS Catalysis} \textbf{\bibinfo{volume}{7}},
	\bibinfo{pages}{8159} (\bibinfo{year}{2017}).
	
	\bibitem[{\citenamefont{Mukherjee et~al.}(2018)\citenamefont{Mukherjee,
			Austeria, and Sampath}}]{Mukherjee2018}
	\bibinfo{author}{\bibfnamefont{D.}~\bibnamefont{Mukherjee}},
	\bibinfo{author}{\bibfnamefont{P.~M.} \bibnamefont{Austeria}},
	\bibnamefont{and} \bibinfo{author}{\bibfnamefont{S.}~\bibnamefont{Sampath}},
	\bibinfo{journal}{ACS Appl. Energy Mater.} \textbf{\bibinfo{volume}{1}},
	\bibinfo{pages}{220} (\bibinfo{year}{2018}).
	
	\bibitem[{\citenamefont{Barua et~al.}(2019)\citenamefont{Barua, Ayyub, Vishnoi,
			Pramoda, and Rao}}]{Barua2019}
	\bibinfo{author}{\bibfnamefont{M.}~\bibnamefont{Barua}},
	\bibinfo{author}{\bibfnamefont{M.~M.} \bibnamefont{Ayyub}},
	\bibinfo{author}{\bibfnamefont{P.}~\bibnamefont{Vishnoi}},
	\bibinfo{author}{\bibfnamefont{K.}~\bibnamefont{Pramoda}}, \bibnamefont{and}
	\bibinfo{author}{\bibfnamefont{C.~N.~R.} \bibnamefont{Rao}},
	\bibinfo{journal}{J. Mater. Chem. A} \textbf{\bibinfo{volume}{7}},
	\bibinfo{pages}{22500} (\bibinfo{year}{2019}).
	
	\bibitem[{\citenamefont{Hao et~al.}(2020)\citenamefont{Hao, Huang, Han, Huang,
			Song, Sun, Wang, Li, Hu, Xue et~al.}}]{Hao2020}
	\bibinfo{author}{\bibfnamefont{Y.}~\bibnamefont{Hao}},
	\bibinfo{author}{\bibfnamefont{A.}~\bibnamefont{Huang}},
	\bibinfo{author}{\bibfnamefont{S.}~\bibnamefont{Han}},
	\bibinfo{author}{\bibfnamefont{H.}~\bibnamefont{Huang}},
	\bibinfo{author}{\bibfnamefont{J.}~\bibnamefont{Song}},
	\bibinfo{author}{\bibfnamefont{X.}~\bibnamefont{Sun}},
	\bibinfo{author}{\bibfnamefont{Z.}~\bibnamefont{Wang}},
	\bibinfo{author}{\bibfnamefont{L.}~\bibnamefont{Li}},
	\bibinfo{author}{\bibfnamefont{F.}~\bibnamefont{Hu}},
	\bibinfo{author}{\bibfnamefont{J.}~\bibnamefont{Xue}}, \bibnamefont{et~al.},
	\bibinfo{journal}{ACS Appl. Mater. Interfaces} \textbf{\bibinfo{volume}{12}},
	\bibinfo{pages}{29393} (\bibinfo{year}{2020}).
	
	\bibitem[{\citenamefont{Yu et~al.}(2019)\citenamefont{Yu, Li, Zhang, Zhou, Li,
			Xu, Zhou, Zhong, and Wang}}]{Yu2019}
	\bibinfo{author}{\bibfnamefont{J.}~\bibnamefont{Yu}},
	\bibinfo{author}{\bibfnamefont{W.-J.} \bibnamefont{Li}},
	\bibinfo{author}{\bibfnamefont{H.}~\bibnamefont{Zhang}},
	\bibinfo{author}{\bibfnamefont{F.}~\bibnamefont{Zhou}},
	\bibinfo{author}{\bibfnamefont{R.}~\bibnamefont{Li}},
	\bibinfo{author}{\bibfnamefont{C.-Y.} \bibnamefont{Xu}},
	\bibinfo{author}{\bibfnamefont{L.}~\bibnamefont{Zhou}},
	\bibinfo{author}{\bibfnamefont{H.}~\bibnamefont{Zhong}}, \bibnamefont{and}
	\bibinfo{author}{\bibfnamefont{J.}~\bibnamefont{Wang}},
	\bibinfo{journal}{Nano Energy} \textbf{\bibinfo{volume}{57}},
	\bibinfo{pages}{222} (\bibinfo{year}{2019}).
	
	\bibitem[{\citenamefont{Brec}(1986)}]{Brec1986}
	\bibinfo{author}{\bibfnamefont{R.}~\bibnamefont{Brec}}, \bibinfo{journal}{Solid
		State Ionics} \textbf{\bibinfo{volume}{22}}, \bibinfo{pages}{3}
	(\bibinfo{year}{1986}).
	
	\bibitem[{\citenamefont{Ismail et~al.}(2010)\citenamefont{Ismail, El-Meligi,
			Temerk, and Madian}}]{Ismail2010}
	\bibinfo{author}{\bibfnamefont{N.}~\bibnamefont{Ismail}},
	\bibinfo{author}{\bibfnamefont{A.}~\bibnamefont{El-Meligi}},
	\bibinfo{author}{\bibfnamefont{Y.}~\bibnamefont{Temerk}}, \bibnamefont{and}
	\bibinfo{author}{\bibfnamefont{M.}~\bibnamefont{Madian}},
	\bibinfo{journal}{Int. J. Hydrog. Energy} \textbf{\bibinfo{volume}{35}},
	\bibinfo{pages}{7827} (\bibinfo{year}{2010}).
	
	\bibitem[{\citenamefont{Evarestov and Kuzmin}(2020)}]{Evarestov2020}
	\bibinfo{author}{\bibfnamefont{R.~A.} \bibnamefont{Evarestov}}
	\bibnamefont{and} \bibinfo{author}{\bibfnamefont{A.}~\bibnamefont{Kuzmin}},
	\bibinfo{journal}{J. Comput. Chem.} \textbf{\bibinfo{volume}{41}},
	\bibinfo{pages}{1337} (\bibinfo{year}{2020}).
	
	\bibitem[{\citenamefont{Dovesi et~al.}(2018)\citenamefont{Dovesi, Erba,
			Orlando, Zicovich-Wilson, Civalleri, Maschio, R\'{e}rat, Casassa, Baima,
			Salustro et~al.}}]{crystal17}
	\bibinfo{author}{\bibfnamefont{R.}~\bibnamefont{Dovesi}},
	\bibinfo{author}{\bibfnamefont{A.}~\bibnamefont{Erba}},
	\bibinfo{author}{\bibfnamefont{R.}~\bibnamefont{Orlando}},
	\bibinfo{author}{\bibfnamefont{C.~M.} \bibnamefont{Zicovich-Wilson}},
	\bibinfo{author}{\bibfnamefont{B.}~\bibnamefont{Civalleri}},
	\bibinfo{author}{\bibfnamefont{L.}~\bibnamefont{Maschio}},
	\bibinfo{author}{\bibfnamefont{M.}~\bibnamefont{R\'{e}rat}},
	\bibinfo{author}{\bibfnamefont{S.}~\bibnamefont{Casassa}},
	\bibinfo{author}{\bibfnamefont{J.}~\bibnamefont{Baima}},
	\bibinfo{author}{\bibfnamefont{S.}~\bibnamefont{Salustro}},
	\bibnamefont{et~al.}, \bibinfo{journal}{WIREs Comput. Mol. Sci.}
	\textbf{\bibinfo{volume}{8}}, \bibinfo{pages}{e1360} (\bibinfo{year}{2018}).
	
	\bibitem[{\citenamefont{Oliveira et~al.}(2019)\citenamefont{Oliveira, Laun,
			Peintinger, and Bredow}}]{Oliveira2019}
	\bibinfo{author}{\bibfnamefont{D.~V.} \bibnamefont{Oliveira}},
	\bibinfo{author}{\bibfnamefont{J.}~\bibnamefont{Laun}},
	\bibinfo{author}{\bibfnamefont{M.~F.} \bibnamefont{Peintinger}},
	\bibnamefont{and} \bibinfo{author}{\bibfnamefont{T.}~\bibnamefont{Bredow}},
	\bibinfo{journal}{J. Comput. Chem.} \textbf{\bibinfo{volume}{40}},
	\bibinfo{pages}{2364} (\bibinfo{year}{2019}).
	
	\bibitem[{\citenamefont{Monkhorst and Pack}(1976)}]{Monkhorst1976}
	\bibinfo{author}{\bibfnamefont{H.~J.} \bibnamefont{Monkhorst}}
	\bibnamefont{and} \bibinfo{author}{\bibfnamefont{J.~D.} \bibnamefont{Pack}},
	\bibinfo{journal}{Phys. Rev. B} \textbf{\bibinfo{volume}{13}},
	\bibinfo{pages}{5188} (\bibinfo{year}{1976}).
	
	\bibitem[{\citenamefont{Becke}(1993)}]{B3LYP}
	\bibinfo{author}{\bibfnamefont{A.~D.} \bibnamefont{Becke}},
	\bibinfo{journal}{J. Chem. Phys.} \textbf{\bibinfo{volume}{98}},
	\bibinfo{pages}{5648} (\bibinfo{year}{1993}).
	
	\bibitem[{\citenamefont{Haines et~al.}(2018)\citenamefont{Haines, Coak, Wildes,
			Lampronti, Liu, Nahai-Williamson, Hamidov, Daisenberger, and
			Saxena}}]{Haines2018}
	\bibinfo{author}{\bibfnamefont{C.~R.~S.} \bibnamefont{Haines}},
	\bibinfo{author}{\bibfnamefont{M.~J.} \bibnamefont{Coak}},
	\bibinfo{author}{\bibfnamefont{A.~R.} \bibnamefont{Wildes}},
	\bibinfo{author}{\bibfnamefont{G.~I.} \bibnamefont{Lampronti}},
	\bibinfo{author}{\bibfnamefont{C.}~\bibnamefont{Liu}},
	\bibinfo{author}{\bibfnamefont{P.}~\bibnamefont{Nahai-Williamson}},
	\bibinfo{author}{\bibfnamefont{H.}~\bibnamefont{Hamidov}},
	\bibinfo{author}{\bibfnamefont{D.}~\bibnamefont{Daisenberger}},
	\bibnamefont{and} \bibinfo{author}{\bibfnamefont{S.~S.}
		\bibnamefont{Saxena}}, \bibinfo{journal}{Phys. Rev. Lett.}
	\textbf{\bibinfo{volume}{121}}, \bibinfo{pages}{266801}
	(\bibinfo{year}{2018}).
	
	\bibitem[{\citenamefont{Jackson et~al.}(2015)\citenamefont{Jackson, Skelton,
			Hendon, Butler, and Walsh}}]{Jackson2015}
	\bibinfo{author}{\bibfnamefont{A.~J.} \bibnamefont{Jackson}},
	\bibinfo{author}{\bibfnamefont{J.~M.} \bibnamefont{Skelton}},
	\bibinfo{author}{\bibfnamefont{C.~H.} \bibnamefont{Hendon}},
	\bibinfo{author}{\bibfnamefont{K.~T.} \bibnamefont{Butler}},
	\bibnamefont{and} \bibinfo{author}{\bibfnamefont{A.}~\bibnamefont{Walsh}},
	\bibinfo{journal}{J. Chem. Phys.} \textbf{\bibinfo{volume}{143}},
	\bibinfo{pages}{184101} (\bibinfo{year}{2015}).
	
	\bibitem[{\citenamefont{Mulliken}(1955)}]{Mulliken1955}
	\bibinfo{author}{\bibfnamefont{R.~S.} \bibnamefont{Mulliken}},
	\bibinfo{journal}{J. Chem. Phys.} \textbf{\bibinfo{volume}{23}},
	\bibinfo{pages}{1833} (\bibinfo{year}{1955}).
	
	\bibitem[{\citenamefont{Aroyo et~al.}(2014)\citenamefont{Aroyo, Orobengoa,
			de~la Flor, Tasci, Perez-Mato, and Wondratschek}}]{Aroyo2014}
	\bibinfo{author}{\bibfnamefont{M.~I.} \bibnamefont{Aroyo}},
	\bibinfo{author}{\bibfnamefont{D.}~\bibnamefont{Orobengoa}},
	\bibinfo{author}{\bibfnamefont{G.}~\bibnamefont{de~la Flor}},
	\bibinfo{author}{\bibfnamefont{E.~S.} \bibnamefont{Tasci}},
	\bibinfo{author}{\bibfnamefont{J.~M.} \bibnamefont{Perez-Mato}},
	\bibnamefont{and}
	\bibinfo{author}{\bibfnamefont{H.}~\bibnamefont{Wondratschek}},
	\bibinfo{journal}{Acta Crystallogr. A} \textbf{\bibinfo{volume}{70}},
	\bibinfo{pages}{126} (\bibinfo{year}{2014}).
	
	\bibitem[{\citenamefont{Pascale et~al.}(2004)\citenamefont{Pascale,
			Zicovich-Wilson, L\'{o}pez~Gejo, Civalleri, Orlando, and
			Dovesi}}]{Pascale2004}
	\bibinfo{author}{\bibfnamefont{F.}~\bibnamefont{Pascale}},
	\bibinfo{author}{\bibfnamefont{C.~M.} \bibnamefont{Zicovich-Wilson}},
	\bibinfo{author}{\bibfnamefont{F.}~\bibnamefont{L\'{o}pez~Gejo}},
	\bibinfo{author}{\bibfnamefont{B.}~\bibnamefont{Civalleri}},
	\bibinfo{author}{\bibfnamefont{R.}~\bibnamefont{Orlando}}, \bibnamefont{and}
	\bibinfo{author}{\bibfnamefont{R.}~\bibnamefont{Dovesi}},
	\bibinfo{journal}{J. Comput. Chem.} \textbf{\bibinfo{volume}{25}},
	\bibinfo{pages}{888} (\bibinfo{year}{2004}).
	
	\bibitem[{\citenamefont{Gatti et~al.}(1994)\citenamefont{Gatti, Saunders, and
			Roetti}}]{Gatti1994}
	\bibinfo{author}{\bibfnamefont{C.}~\bibnamefont{Gatti}},
	\bibinfo{author}{\bibfnamefont{V.~R.} \bibnamefont{Saunders}},
	\bibnamefont{and} \bibinfo{author}{\bibfnamefont{C.}~\bibnamefont{Roetti}},
	\bibinfo{journal}{J. Chem. Phys.} \textbf{\bibinfo{volume}{101}},
	\bibinfo{pages}{10686} (\bibinfo{year}{1994}).
	
	\bibitem[{\citenamefont{Casassa et~al.}(2015)\citenamefont{Casassa, Erba,
			Baima, and Orlando}}]{Casassa2015}
	\bibinfo{author}{\bibfnamefont{S.}~\bibnamefont{Casassa}},
	\bibinfo{author}{\bibfnamefont{A.}~\bibnamefont{Erba}},
	\bibinfo{author}{\bibfnamefont{J.}~\bibnamefont{Baima}}, \bibnamefont{and}
	\bibinfo{author}{\bibfnamefont{R.}~\bibnamefont{Orlando}},
	\bibinfo{journal}{J. Comput. Chem.} \textbf{\bibinfo{volume}{36}},
	\bibinfo{pages}{1940} (\bibinfo{year}{2015}).
	
	\bibitem[{\citenamefont{Adamo and Barone}(1999)}]{PBE0}
	\bibinfo{author}{\bibfnamefont{C.}~\bibnamefont{Adamo}} \bibnamefont{and}
	\bibinfo{author}{\bibfnamefont{V.}~\bibnamefont{Barone}},
	\bibinfo{journal}{J. Chem. Phys.} \textbf{\bibinfo{volume}{110}},
	\bibinfo{pages}{6158} (\bibinfo{year}{1999}).
	
	\bibitem[{\citenamefont{Zhao and Truhlar}(2008)}]{M06}
	\bibinfo{author}{\bibfnamefont{Y.}~\bibnamefont{Zhao}} \bibnamefont{and}
	\bibinfo{author}{\bibfnamefont{D.~G.} \bibnamefont{Truhlar}},
	\bibinfo{journal}{Theor. Chem. Account.} \textbf{\bibinfo{volume}{120}},
	\bibinfo{pages}{215} (\bibinfo{year}{2008}).
	
	\bibitem[{\citenamefont{Du et~al.}(2016)\citenamefont{Du, Wang, Liu, Hu, Utama,
			Gan, Xiong, and Kloc}}]{Du2016}
	\bibinfo{author}{\bibfnamefont{K.-Z.} \bibnamefont{Du}},
	\bibinfo{author}{\bibfnamefont{X.-Z.} \bibnamefont{Wang}},
	\bibinfo{author}{\bibfnamefont{Y.}~\bibnamefont{Liu}},
	\bibinfo{author}{\bibfnamefont{P.}~\bibnamefont{Hu}},
	\bibinfo{author}{\bibfnamefont{M.~I.~B.} \bibnamefont{Utama}},
	\bibinfo{author}{\bibfnamefont{C.~K.} \bibnamefont{Gan}},
	\bibinfo{author}{\bibfnamefont{Q.}~\bibnamefont{Xiong}}, \bibnamefont{and}
	\bibinfo{author}{\bibfnamefont{C.}~\bibnamefont{Kloc}}, \bibinfo{journal}{ACS
		Nano} \textbf{\bibinfo{volume}{10}}, \bibinfo{pages}{1738}
	(\bibinfo{year}{2016}).
	
	\bibitem[{\citenamefont{Baroni et~al.}(2001)\citenamefont{Baroni, de~Gironcoli,
			Dal~Corso, and Giannozzi}}]{Baroni2001}
	\bibinfo{author}{\bibfnamefont{S.}~\bibnamefont{Baroni}},
	\bibinfo{author}{\bibfnamefont{S.}~\bibnamefont{de~Gironcoli}},
	\bibinfo{author}{\bibfnamefont{A.}~\bibnamefont{Dal~Corso}},
	\bibnamefont{and}
	\bibinfo{author}{\bibfnamefont{P.}~\bibnamefont{Giannozzi}},
	\bibinfo{journal}{Rev. Mod. Phys.} \textbf{\bibinfo{volume}{73}},
	\bibinfo{pages}{515} (\bibinfo{year}{2001}).
	
	\bibitem[{\citenamefont{Scagliotti et~al.}(1987)\citenamefont{Scagliotti,
			Jouanne, Balkanski, Ouvrard, and Benedek}}]{Scagliotti1987}
	\bibinfo{author}{\bibfnamefont{M.}~\bibnamefont{Scagliotti}},
	\bibinfo{author}{\bibfnamefont{M.}~\bibnamefont{Jouanne}},
	\bibinfo{author}{\bibfnamefont{M.}~\bibnamefont{Balkanski}},
	\bibinfo{author}{\bibfnamefont{G.}~\bibnamefont{Ouvrard}}, \bibnamefont{and}
	\bibinfo{author}{\bibfnamefont{G.}~\bibnamefont{Benedek}},
	\bibinfo{journal}{Phys. Rev. B} \textbf{\bibinfo{volume}{35}},
	\bibinfo{pages}{7097} (\bibinfo{year}{1987}).
	
	\bibitem[{\citenamefont{Bernasconi et~al.}(1988)\citenamefont{Bernasconi,
			Marra, Benedek, Miglio, Jouanne, Julien, Scagliotti, and
			Balkanski}}]{Bernasconi1988}
	\bibinfo{author}{\bibfnamefont{M.}~\bibnamefont{Bernasconi}},
	\bibinfo{author}{\bibfnamefont{G.~L.} \bibnamefont{Marra}},
	\bibinfo{author}{\bibfnamefont{G.}~\bibnamefont{Benedek}},
	\bibinfo{author}{\bibfnamefont{L.}~\bibnamefont{Miglio}},
	\bibinfo{author}{\bibfnamefont{M.}~\bibnamefont{Jouanne}},
	\bibinfo{author}{\bibfnamefont{C.}~\bibnamefont{Julien}},
	\bibinfo{author}{\bibfnamefont{M.}~\bibnamefont{Scagliotti}},
	\bibnamefont{and}
	\bibinfo{author}{\bibfnamefont{M.}~\bibnamefont{Balkanski}},
	\bibinfo{journal}{Phys. Rev. B} \textbf{\bibinfo{volume}{38}},
	\bibinfo{pages}{12089} (\bibinfo{year}{1988}).
	
	\bibitem[{\citenamefont{Kroumova et~al.}(2003)\citenamefont{Kroumova, Aroyo,
			Perez-Mato, Kirov, Capillas, Ivantchev, and Wondratschek}}]{Kroumova2003}
	\bibinfo{author}{\bibfnamefont{E.}~\bibnamefont{Kroumova}},
	\bibinfo{author}{\bibfnamefont{M.}~\bibnamefont{Aroyo}},
	\bibinfo{author}{\bibfnamefont{J.}~\bibnamefont{Perez-Mato}},
	\bibinfo{author}{\bibfnamefont{A.}~\bibnamefont{Kirov}},
	\bibinfo{author}{\bibfnamefont{C.}~\bibnamefont{Capillas}},
	\bibinfo{author}{\bibfnamefont{S.}~\bibnamefont{Ivantchev}},
	\bibnamefont{and}
	\bibinfo{author}{\bibfnamefont{H.}~\bibnamefont{Wondratschek}},
	\bibinfo{journal}{Phase Transit.} \textbf{\bibinfo{volume}{76}},
	\bibinfo{pages}{155} (\bibinfo{year}{2003}).
	
	\bibitem[{\citenamefont{Lua\~{n}a et~al.}(2007)\citenamefont{Lua\~{n}a, Blanco,
			Costales, Mari-S\'{a}nchez, and Pend\'{a}s}}]{Luana2007}
	\bibinfo{author}{\bibfnamefont{V.}~\bibnamefont{Lua\~{n}a}},
	\bibinfo{author}{\bibfnamefont{M.~A.} \bibnamefont{Blanco}},
	\bibinfo{author}{\bibfnamefont{A.}~\bibnamefont{Costales}},
	\bibinfo{author}{\bibfnamefont{P.}~\bibnamefont{Mari-S\'{a}nchez}},
	\bibnamefont{and} \bibinfo{author}{\bibfnamefont{A.~M.}
		\bibnamefont{Pend\'{a}s}}, in \emph{\bibinfo{booktitle}{The Quantum Theory of
			Atoms in Molecules: From Solid State to DNA and Drug Design}}, edited by
	\bibinfo{editor}{\bibfnamefont{C.~F.} \bibnamefont{Matta}} \bibnamefont{and}
	\bibinfo{editor}{\bibfnamefont{R.~J.} \bibnamefont{Boyd}}
	(\bibinfo{publisher}{Wiley-VCH Verlag GmBH}, \bibinfo{address}{Darmstadt},
	\bibinfo{year}{2007}), chap.~\bibinfo{chapter}{8}, pp.
	\bibinfo{pages}{207--229}.
	
	\bibitem[{\citenamefont{Morse and Cairns}(1969)}]{Morse1969}
	\bibinfo{author}{\bibfnamefont{M.}~\bibnamefont{Morse}} \bibnamefont{and}
	\bibinfo{author}{\bibfnamefont{S.~S.} \bibnamefont{Cairns}},
	\emph{\bibinfo{title}{Critical Point Theory in Global Analysis and
			Differential Geometry}} (\bibinfo{publisher}{Academic Press},
	\bibinfo{address}{New York}, \bibinfo{year}{1969}).
	
	\bibitem[{\citenamefont{Cremer and Kraka}(1984)}]{Cremer1984}
	\bibinfo{author}{\bibfnamefont{D.}~\bibnamefont{Cremer}} \bibnamefont{and}
	\bibinfo{author}{\bibfnamefont{E.}~\bibnamefont{Kraka}},
	\bibinfo{journal}{Croat. Chem. Acta.} \textbf{\bibinfo{volume}{57}},
	\bibinfo{pages}{1259} (\bibinfo{year}{1984}).
	
	\bibitem[{\citenamefont{Espinosa et~al.}(2002)\citenamefont{Espinosa, Alkorta,
			Elguero, and Molins}}]{Espinosa2002}
	\bibinfo{author}{\bibfnamefont{E.}~\bibnamefont{Espinosa}},
	\bibinfo{author}{\bibfnamefont{I.}~\bibnamefont{Alkorta}},
	\bibinfo{author}{\bibfnamefont{J.}~\bibnamefont{Elguero}}, \bibnamefont{and}
	\bibinfo{author}{\bibfnamefont{E.}~\bibnamefont{Molins}},
	\bibinfo{journal}{J. Chem. Phys.} \textbf{\bibinfo{volume}{117}},
	\bibinfo{pages}{5529} (\bibinfo{year}{2002}).
	
	\bibitem[{\citenamefont{Marabello et~al.}(2004)\citenamefont{Marabello,
			Bianchi, Gervasio, and Cargnoni}}]{Marabello2004}
	\bibinfo{author}{\bibfnamefont{D.}~\bibnamefont{Marabello}},
	\bibinfo{author}{\bibfnamefont{R.}~\bibnamefont{Bianchi}},
	\bibinfo{author}{\bibfnamefont{G.}~\bibnamefont{Gervasio}}, \bibnamefont{and}
	\bibinfo{author}{\bibfnamefont{F.}~\bibnamefont{Cargnoni}},
	\bibinfo{journal}{Acta Crystallogr. A} \textbf{\bibinfo{volume}{60}},
	\bibinfo{pages}{494} (\bibinfo{year}{2004}).
	
	\bibitem[{\citenamefont{Tsirelson}(2014)}]{Tsirelson2014}
	\bibinfo{author}{\bibfnamefont{V.~G.} \bibnamefont{Tsirelson}},
	\emph{\bibinfo{title}{Quantum Chemistry. Molecules Molecular Systems and
			Solids}} (\bibinfo{publisher}{Binom Publ.}, \bibinfo{address}{Moscow},
	\bibinfo{year}{2014}).
	
	\bibitem[{\citenamefont{Momma and Izumi}(2011)}]{VESTA}
	\bibinfo{author}{\bibfnamefont{K.}~\bibnamefont{Momma}} \bibnamefont{and}
	\bibinfo{author}{\bibfnamefont{F.}~\bibnamefont{Izumi}}, \bibinfo{journal}{J.
		Appl. Crystallogr.} \textbf{\bibinfo{volume}{44}}, \bibinfo{pages}{1272}
	(\bibinfo{year}{2011}).
	
\end{thebibliography}


\clearpage

\begin{figure}
	\centering
	\caption{ Crystallographic structure  of FePSe$_3$ in the low-pressure (LP, $P$=0 GPa) trigonal (space group $R\bar{3}$) phase, intermediate pressure (HP-I, $P$=10 GPa) monoclinic (space group $C2/m$) phase and high-pressure (HP-II, $P$=20 GPa) trigonal (space group $P\bar{3}1m$) phase. Conventional unit cells are shown. The distance between the two adjacent layers is denoted as $d$. The illustrations were created using the VESTA software.\protect\cite{VESTA}}
	\label{fig1}
\end{figure}

\begin{figure}
	\centering
	\caption{Pressure dependence of the calculated lattice parameters and the band gap  $E_g$ in FePSe$_3$. Open circles in the lower panel show extrapolation of the band gap beyond the HP-I and HP-II phase existence ranges.  }
	\label{fig2}
\end{figure}

\begin{figure*}
	\centering
	\caption{Band structure diagram for the LP, HP-I, and HP-II FePSe$_3$ phases.
		The energy zero is set at the top of the valence band (Fermi energy position).  }
	\label{fig3}
\end{figure*}

\begin{figure*}
	\centering
	\caption{Total and projected density of states (DOS) for the LP, HP-I, and HP-II FePSe$_3$ phases. The energy zero is set at the top of the valence band (Fermi energy position). }
	\label{fig4}
\end{figure*}

\begin{figure}
	\centering
	\caption{Dependence of the band gap $E_g$ in LP, HP-I, and HP-II FePSe$_3$ phases on the displacement of phosphorus atoms $\Delta z$(P) along the $c$-axis.  }
	\label{fig5}
\end{figure}

\begin{figure*}
	\centering
	\caption{Total and projected density of states (DOS) for the HP-I ($C2/m$) FePSe$_3$ phase as a function of phosphorus atoms displacement $\Delta z$(P) along the $c$-axis. The energy zero is set at the top of the valence band (Fermi energy position). }
	\label{fig6}
\end{figure*}

\begin{figure*}
	\centering
	\caption{The electron localization function (ELF) contour maps for LP (0 GPa), HP-I (10 GPa), and HP-II (20 GPa) phases of FePSe$_3$. Selected atoms are indicated. Atom numbering is as in Fig.\ \protect\ref{fig1}.}
	\label{fig7}
\end{figure*}


\clearpage

\begin{center}
	\includegraphics[width=0.7\columnwidth,keepaspectratio=true]{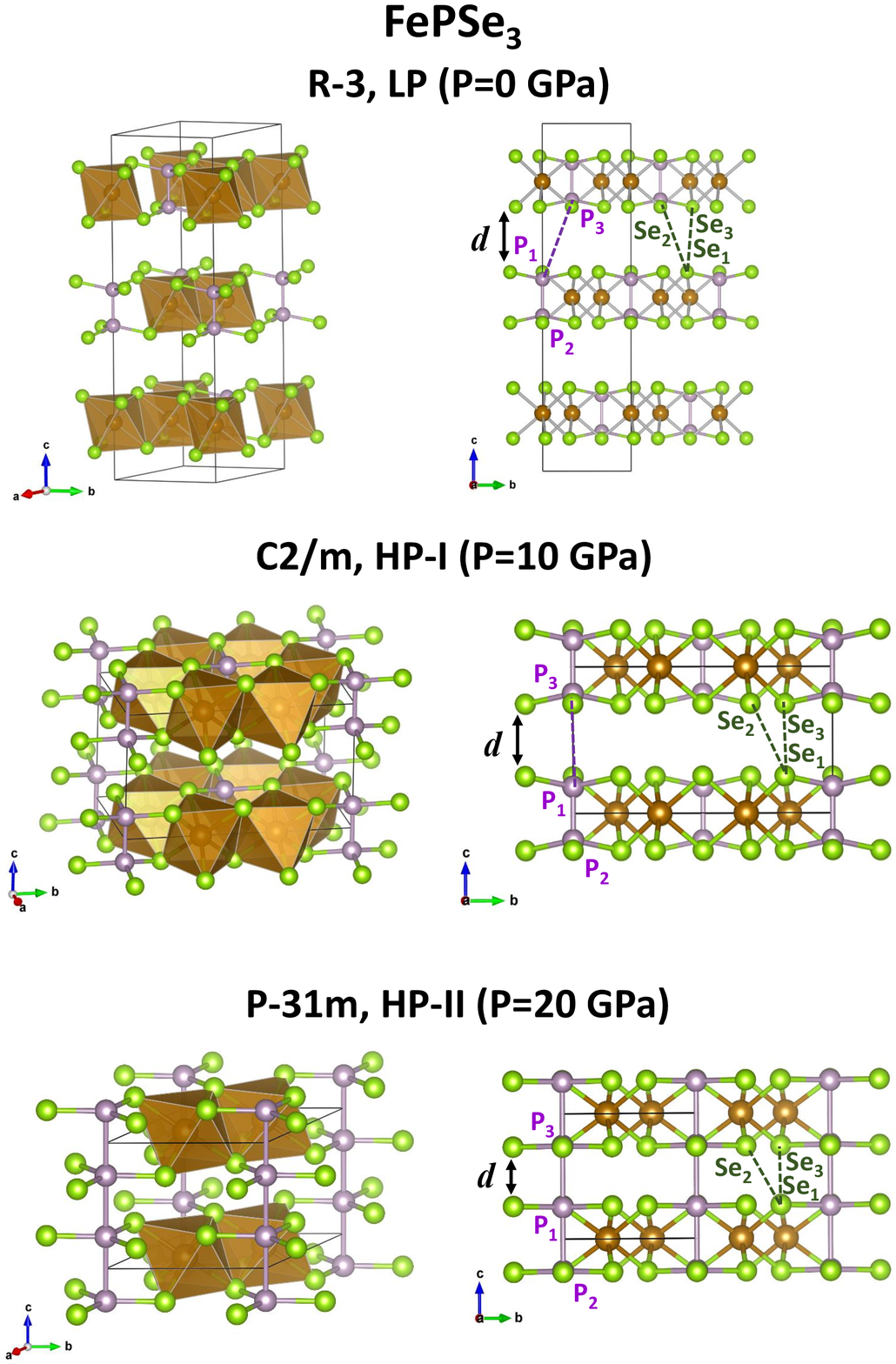}
\end{center}
\vspace{0.25in}
\hspace*{3in}
{\Large
	\begin{minipage}[t]{3in}
		\baselineskip = .5\baselineskip
		Figure 1 \\
		R. A. Evarestov, A. Kuzmin \\
		J.\ Comput.\ Chem.
	\end{minipage}
}

\clearpage

\begin{center}
	\includegraphics[width=0.5\columnwidth,keepaspectratio=true]{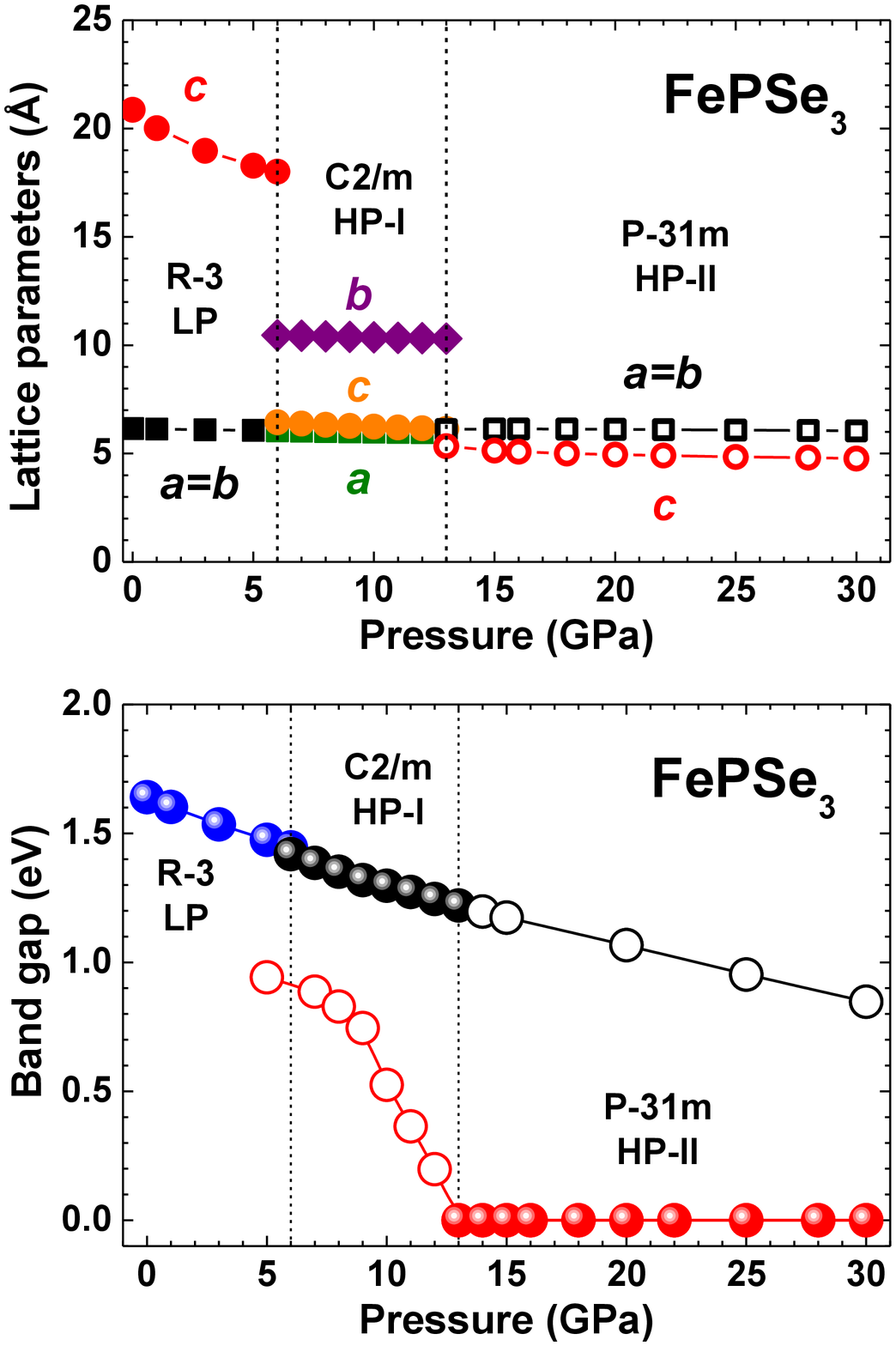}
\end{center}
\vspace{0.25in}
\hspace*{3in}
{\Large
	\begin{minipage}[t]{3in}
		\baselineskip = .5\baselineskip
		Figure 2 \\
		R. A. Evarestov, A. Kuzmin \\
		J.\ Comput.\ Chem.
	\end{minipage}
}

\clearpage

\begin{center}
	\includegraphics[width=0.9\columnwidth,keepaspectratio=true]{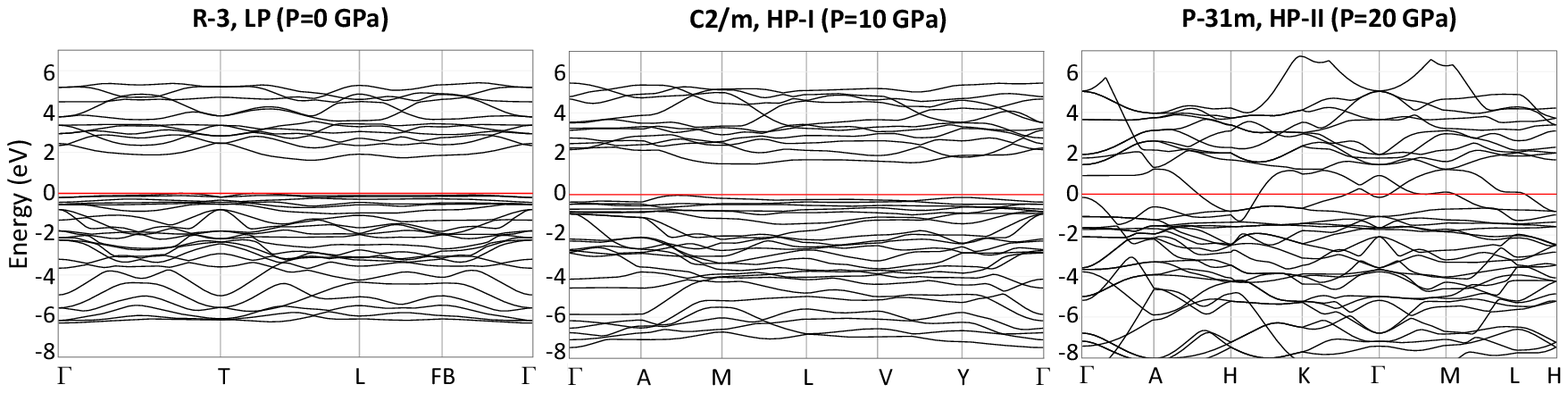}
\end{center}
\vspace{0.25in}
\hspace*{3in}
{\Large
	\begin{minipage}[t]{3in}
		\baselineskip = .5\baselineskip
		Figure 3 \\
		R. A. Evarestov, A. Kuzmin \\
		J.\ Comput.\ Chem.
	\end{minipage}
}

\clearpage

\begin{center}
	\includegraphics[width=0.9\columnwidth,keepaspectratio=true]{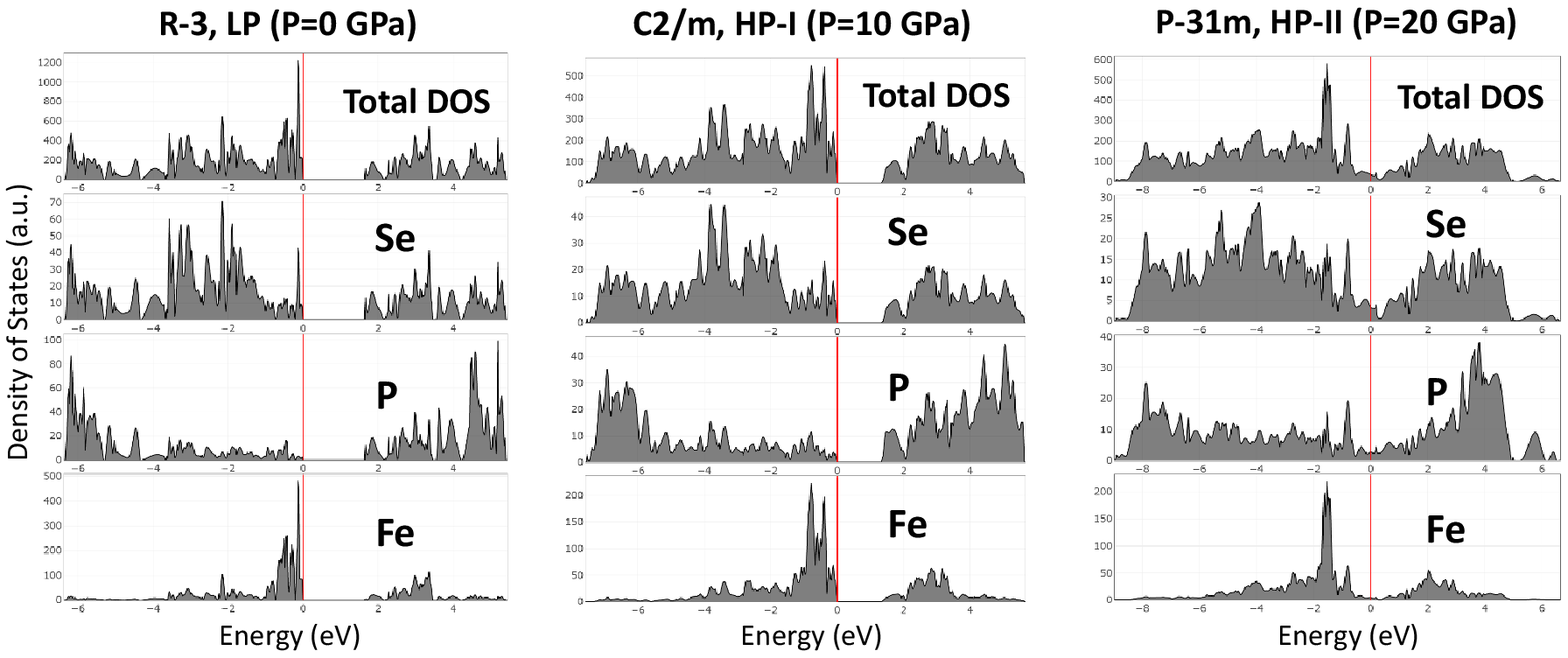}
\end{center}
\vspace{0.25in}
\hspace*{3in}
{\Large
	\begin{minipage}[t]{3in}
		\baselineskip = .5\baselineskip
		Figure 4 \\
		R. A. Evarestov, A. Kuzmin \\
		J.\ Comput.\ Chem.
	\end{minipage}
}

\clearpage

\begin{center}
	\includegraphics[width=0.5\columnwidth,keepaspectratio=true]{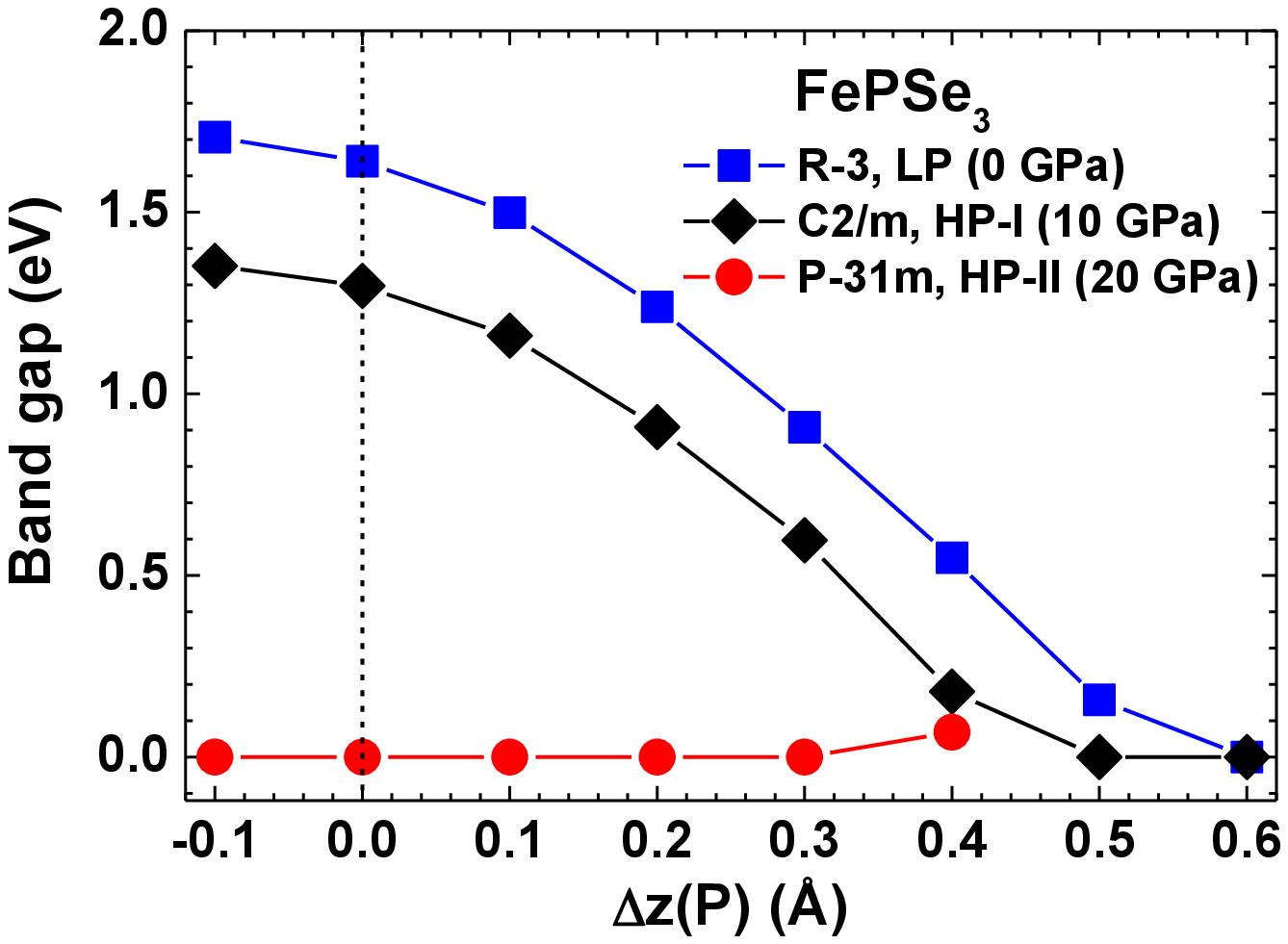}
\end{center}
\vspace{0.25in}
\hspace*{3in}
{\Large
	\begin{minipage}[t]{3in}
		\baselineskip = .5\baselineskip
		Figure 5 \\
		R. A. Evarestov, A. Kuzmin \\
		J.\ Comput.\ Chem.
	\end{minipage}
}

\clearpage

\begin{center}
	\includegraphics[width=0.8\columnwidth,keepaspectratio=true]{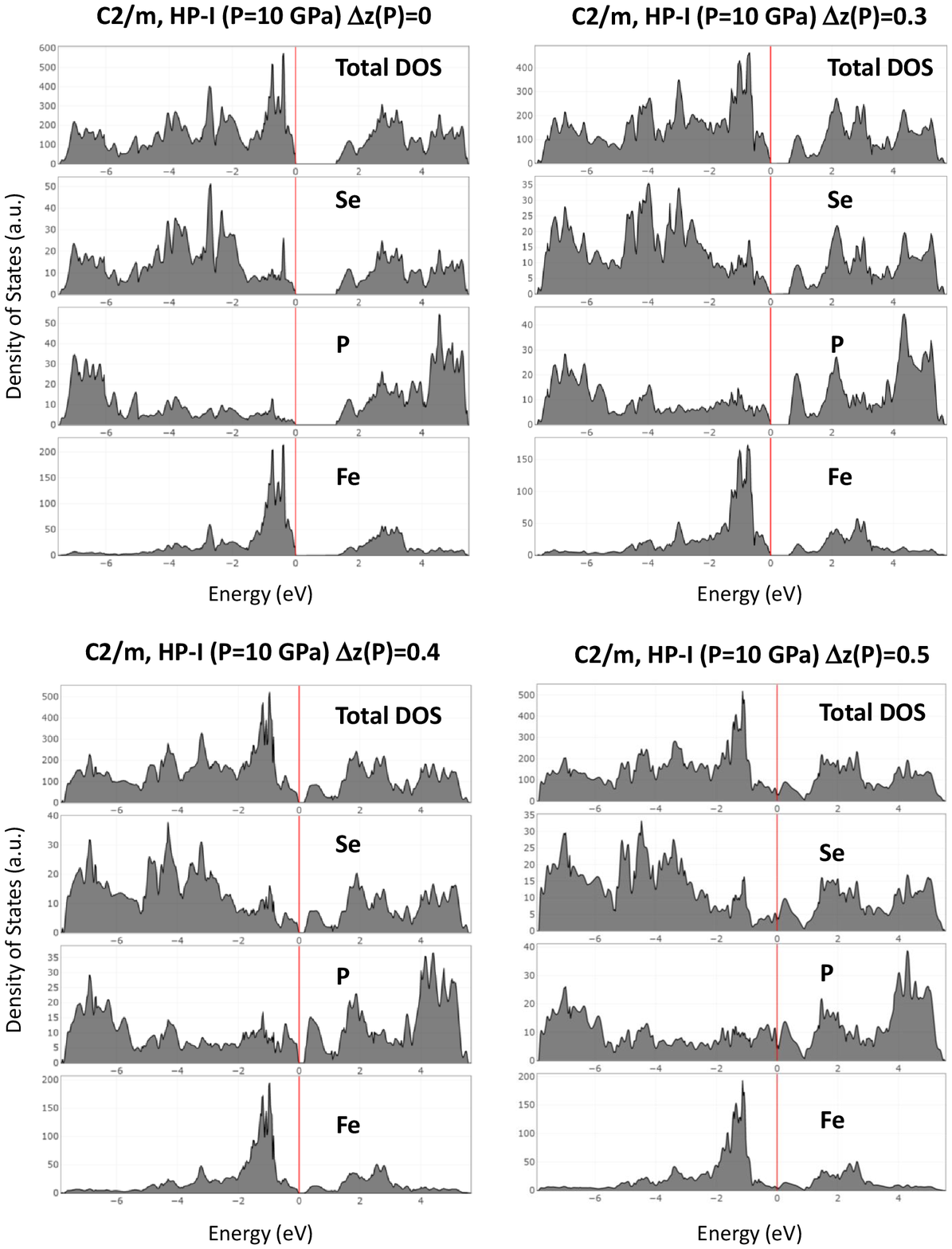}
\end{center}
\vspace{0.25in}
\hspace*{3in}
{\Large
	\begin{minipage}[t]{3in}
		\baselineskip = .5\baselineskip
		Figure 6 \\
		R. A. Evarestov, A. Kuzmin \\
		J.\ Comput.\ Chem.
	\end{minipage}
}

\clearpage

\begin{center}
	\includegraphics[width=0.9\columnwidth,keepaspectratio=true]{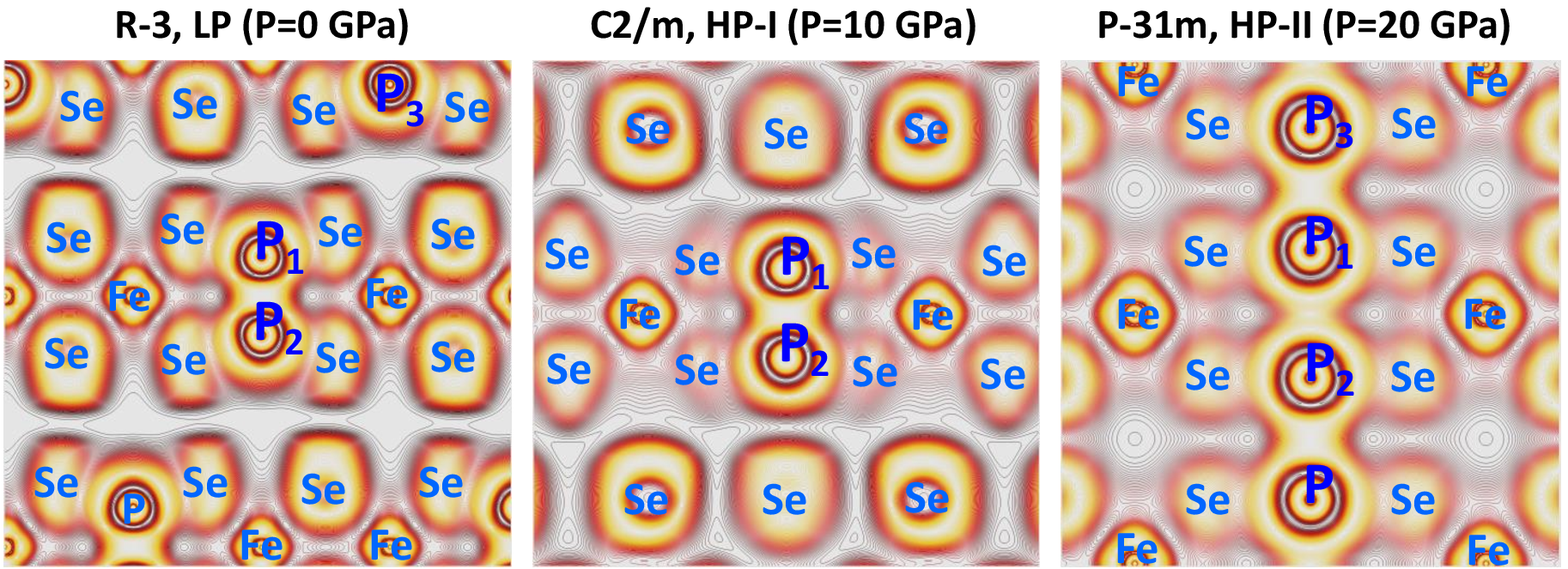}
\end{center}
\vspace{0.25in}
\hspace*{3in}
{\Large
	\begin{minipage}[t]{3in}
		\baselineskip = .5\baselineskip
		Figure 7 \\
		R. A. Evarestov, A. Kuzmin \\
		J.\ Comput.\ Chem.
	\end{minipage}
}

\clearpage


\begin{table*}
	\footnotesize
	\caption{Crystallographic parameters, band gap values and atomic charges for FePSe$_3$ at 0, 10, and 20 GPa. Experimental data are taken from Refs. \protect\cite{WIEDENMANN1981,Du2016}.  }
	\label{table1}     
	\centering 
	\renewcommand{\arraystretch}{0.8}
	\begin{tabular}{lllcc} 
		\\
		\hline 
		& \multicolumn{2}{c}{Space group $R\bar{3}$ (148)}  &  Space group $C2/m$ (12) & Space group $P\bar{3}1m$ (162)  \\  
		& \multicolumn{2}{c}{LP ($P$=0 GPa)}   & HP-I ($P$=10 GPa) &  HP-II ($P$=20 GPa)  \\ 
		
		& Experiment \protect\cite{WIEDENMANN1981} & LCAO     & LCAO      & LCAO  \\        
		\hline
		a (\AA)  & 6.262  &  6.154 &     5.989   & 6.118  \\
		b (\AA)  &  &   &     10.37     &       \\
		c (\AA)  & 19.81 & 20.87     & 6.244    & 4.953   \\
		
		$\beta$ ($^\circ$) &  &    & 108.8  & \\
		
		y(Fe)   &        &         & 0.3333     &  \\
		z(Fe)   & 0.1667 & 0.1665  &      &  \\
		
		x(P)   &         &          & 0.0628    &   \\		
		z(P)   & 0.4486  & 0.4477   & 0.1824    &  -0.2565 		 \\
		
		x(Se1)  & 0.3282	 & 0.3340    &   -0.2836 &  0.3700	 \\
		y(Se1)  & 0.0073	 & 0.0320    &           &   \\
		z(Se1)  & 0.0801 	 & 0.0925    & 0.2513    &   -0.2637  \\		

		x(Se2)  &     		 &       & 0.2686 		  &   \\
		y(Se2)  &     		 &       & 0.1849 		   &   \\	
		z(Se2)  &      	     &       & 0.2492 		  &    \\

		$E_g$ (eV) & 1.3 \protect\cite{Du2016}      &  1.6 &   1.3 & 0 \\ 
		\\
                &  \multicolumn{4}{c}{\em Bader charges} \\ 
 		Fe      &        &   0.52  &   0.40    &   0.26  \\       
 		P       &        &   0.19  &   0.19    &   0.12  \\   		
 		Se1     &        &  -0.23  &  -0.22    &  -0.13  \\  
 		Se2     &        &         &  -0.18    &         \\   		 		 
                               
                  \\
                & \multicolumn{4}{c}{\em Mulliken charges} \\ 
 		Fe      &        &  0.29   &  0.16    &  -0.13    \\       
        P       &        &  0.60   &  0.62    &   0.47    \\   		
        Se1     &        & -0.30   & -0.30    &  -0.11    \\  
        Se2     &        &         & -0.24    &           \\                 

		\hline
	\end{tabular}
\end{table*}   

\clearpage

\begin{table*}
	\footnotesize
	\caption{Calculated phonon frequencies (in cm$^{-1}$) at the  $\Gamma$-point for FePSe$_3$ at $P$=0, 10, and 20 GPa. Acoustic (A), Raman active (R), infrared active (IR), and silent (S) modes are indicated. The experimental infrared and Raman frequencies (exp.) measured at room temperature \protect\cite{Scagliotti1987,Bernasconi1988} are also reported for comparison. }
	\label{table2}     
	\centering 
	\renewcommand{\arraystretch}{0.8}
	\begin{tabular}{llllllllll} 
		\\
		\hline
		\multicolumn{4}{c}{Space group $R\bar{3}$ (148)}  & \multicolumn{3}{c}{Space group $C2/m$ (12)}  & \multicolumn{3}{c}{Space group $P\bar{3}1m$ (162)}  \\  
		\multicolumn{4}{c}{LP ($P$=0 GPa)}   &   \multicolumn{3}{c}{HP-I ($P$=10 GPa)}  & \multicolumn{3}{c}{HP-II ($P$=20 GPa)}  \\ 
		Mode & Frequency &  Frequency (exp.)& Activity  & Mode & Frequency & Activity  &  Mode & Frequency & Activity   \\    
		\hline
E$_u$ &	0  &    &	A	&	A$_u$   &	 0	&	A	&	A$_{2u}$ &		    0	&	A        \\ 	
A$_u$ &	0  &    &	A	&	B$_u$   &	 0	&	A	&	E$_u$ 	  &	      0	&	A    \\ 	
A$_g$ &	111&  74	&	R	&	B$_u$ 	&	 0	&	IR	&	A$_{2u}$ &		122	&	IR        \\ 	
A$_u$ &	118& 108	&	IR	&	B$_u$   &   110	&	IR	&	A$_{2g}$  &	 	150	&	S   \\ 	
E$_g$ &	122& 117	&	R	&	B$_g$ 	&	121	&	R	&	E$_u$  	  &	  160	&	IR           \\	 
E$_u$ &	131& 126	&	IR	&	B$_g$ 	&	140	&	R	&	E$_g$  	  &	  174	&	R         \\ 	
E$_g$ &	154& 139	&	R	&	A$_g$ 	&	141	&	R	&	A$_{1g}$  &	 	180	&	R         \\ 	
A$_g$ &	158& 150	&	R	&	B$_u$ 	&	158	&	IR	&	E$_g$	      &	  194	&	R         \\ 	
E$_u$ &	160& 159	&	IR	&	A$_u$ 	&	161	&	IR	&	E$_u$ 	  &	  195	&	IR        \\ 	
A$_u$ &	165& 	&	IR	&	B$_u$ 	&	167	&	IR	&	A$_{2u}$ &		197	&	IR        \\ 	
E$_g$ &	178& 172	&	R	&	B$_g$ 	&	172	&	R	&	E$_g$ 	  &	  212	&	R         \\ 	
A$_g$ &	202& 218	&	R	&	A$_g$ 	&	172	&	R	&	A$_{1u}$ &		238	&	S          \\	 
A$_u$ &	212&	&	IR	&	A$_u$ 	&	177	&	IR	&	A$_{1g}$  &	 	243	&	R         \\ 	
A$_g$ &	230&  	&	R	&	A$_g$ 	&	193	&	R	&	A$_{1g}$  &	 	263	&	R         \\ 	
E$_u$ &	257&	&	IR	&	B$_g$ 	&	195	&	R	&	A$_{2u}$ &		320	&	IR        \\ 	
E$_g$ &	267& 	&	R	&	A$_u$ 	&	201	&	IR	&	A$_{2g}$  &	 	321	&	S          \\	 
A$_u$ &	288& 305	&	IR	&	A$_g$ 	&	202	&	R	&	E$_u$    &	331	&	IR        \\ 	
E$_u$ &	393& 442	&	IR	&	A$_g$ 	&	227	&	R	&	E$_g$ 	  &	  333	&	R         \\ 	
E$_g$ &	395& 450	&	R	&	B$_u$ 	&	243	&	IR	&	E$_g$ 	  &	  414	&	R         \\ 	
A$_g$ &	516& 506	&	R	&	B$_g$ 	&	279	&	R	&	E$_u$ 	  &	  432	&	IR        \\ 	
 & & & & Au 	& 295	& IR  &     &    &     \\
 & & & & Bu 	& 295	& IR  &     &    &      \\
 & & & & Bu 	& 301	& IR  &     &    &     \\
 & & & & Ag 	& 308	& R   &     &    &     \\
 & & & & Bg 	& 309	& R   &     &    &     \\
 & & & & Ag 	& 428	& R   &     &    &     \\
 & & & & Bu 	& 430	& IR  &     &    &      \\
 & & & & Bg 	& 430	& R   &     &    &     \\
 & & & & Au 	& 434	& IR  &     &    &     \\
 & & & & Ag 	& 530	& R   &     &    &     \\ 
		\hline
	\end{tabular}
\end{table*}

\clearpage

\begin{table*}
	\footnotesize
	\caption{Critical points and their Wyckoff positions (WP) in the primitive cell of FePSe$_3$. 
		The Morse relationships $n-b+r-c=0$ as expected in crystalline materials  ($30-87+99-42=0$ for LP, $10-25+28-13=0$ for HP-I and $10-29+27-8=0$ for HP-II).  }
	\label{table3}     
	\centering 
	\renewcommand{\arraystretch}{0.8}
	\begin{tabular}{crllcrllcrll} 
		\\
		\hline 
		\multicolumn{4}{c}{LP ($R\bar{3}$)} & 		\multicolumn{4}{c}{HP-I ($C2/m$)}  & \multicolumn{4}{c}{HP-II ($P\bar{3}1m$)} \\
		CP No.  &  \multicolumn{2}{c}{CP type} &  WP  &   CP No.  &  \multicolumn{2}{c}{CP type} &  WP  &   CP No.  &  \multicolumn{2}{c}{CP type} &  WP \\
		\hline
		1	 & (3,-3)  &   n  & 6c (Fe)     &   1	 & (3,-3)    & n    & 2g (Fe)	 &   1	 &  (3,-3)  &  n   & 2c	(Fe) \\  
		2	 & (3,-3)  &   n  & 6c  (P)   &   2	 & (3,-3)    & n    & 2i (P)	 &   2	 &  (3,-3)  &  n   & 2e	(P) \\  
		3	 & (3,-3)  &   n  & 18f (Se)   &   3	 & (3,-3)    & n    & 2i (Se1)	 &   3	 &  (3,-3)  &  n   & 6k (Se)	 \\  
		4	 & (3,-1)  &   b  & 18f    &   4	 & (3,-3)    & n    & 4j (Se2)	 &   4	 &  (3,-1)  &  b   & 6k	 \\  
		5	 & (3,-1)  &   b  & 18f    &   5	 & (3,-1)    & b    & 4j	 &   5	 &  (3,-1)  &  b   & 12l	 \\  
		6	 & (3,-1)  &   b  & 9e     &   6	 & (3,-1)    & b    & 4j	 &   6	 &  (3,-1)  &  b   & 6j	 \\  
		7	 & (3,-1)  &   b  & 6c     &   7	 & (3,-1)    & b    & 4j	 &   7	 &  (3,-1)  &  b   & 3g	 \\  
		8	 & (3,-1)  &   b  & 18f    &   8     & (3,-1)    & b    & 1a	 &   8	 &  (3,-1)  &  b   & 1a	 \\  
		9	 & (3,-1)  &   b  & 18f    &   9	 & (3,-1)    & b    & 4j	 &   9	 &  (3,-1)  &  b   & 1b	 \\  
		10	 & (3,+1)  &   r  & 9d     &   10	 & (3,-1)    & b    & 2i	 &   10	 &  (3,+1)  &  r   & 3f	 \\  
		11	 & (3,+1)  &   r  & 18f    &   11	 & (3,-1)    & b    & 2i	 &   11	 &  (3,+1)  &  r   & 12l	 \\  
		12	 & (3,+1)  &   r  & 18f    &   12	 & (3,-1)    & b    & 2f	 &   12	 &  (3,+1)  &  r   & 6i	 \\  
		13	 & (3,+1)  &   r  & 18f    &   13	 & (3,-1)    & b    & 2h	 &   13	 &  (3,+1)  &  r   & 6j	 \\  
		14	 & (3,+1)  &   r  & 18f    &   14	 & (3,+1)    & r    & 4j	 &   14	 &  (3,+3)  &  c   & 6k	 \\  
		15	 & (3,+1)  &   r  & 18f    &   15	 & (3,+1)    & r    & 1d	 &   15	 &  (3,+3)  &  c   & 2d	 \\  
		16	 & (3,+3)  &   c  & 6c     &   16	 & (3,+1)    & r    & 1c	 &       &          &      &  \\  
		17	 & (3,+3)  &   c  & 18f    &   17	 & (3,+1)    & r    & 1b	 &       &          &      &  \\  
		18	 & (3,+3)  &   c  & 18f    &   18	 & (3,+1)    & r    & 4j	 &       &          &      &  \\  
		&         &      &        &   19	 & (3,+1)    & r    & 4j	 &       &          &      &  \\  
		& 	       &      &        &   20	 & (3,+1)    & r    & 2i	 &       &          &      &  \\  
		& 	       &      &        &   21	 & (3,+1)    & r    & 4j	 &       &          &      &  \\  
		& 	       &      &        &   22	 & (3,+1)    & r    & 4j	 &       &          &      &  \\  
		& 	       &      &        &   23	 & (3,+1)    & r    & 2g	 &       &          &      &  \\  
		& 	       &      &        &   24	 & (3,+3)    & c	& 2e     &       &          &      & \\  
		& 	       &      &        &   25	 & (3,+3)    & c    & 2h	 &       &          &      &  \\  
		& 	       &      &        &   26	 & (3,+3)    & c    & 4j	 &       &          &      &  \\  
		& 	       &      &        &   27	 & (3,+3)    & c    & 2i	 &       &          &      &  \\  
		& 	       &      &        &   28	 & (3,+3)    & c    & 4j	 &       &          &      &  \\  
		\hline
	\end{tabular}
\end{table*}

\begin{table*}
	\footnotesize
	\caption{Results of the topological analysis of the electron density ($\rho$) and its Laplacian ($\bigtriangledown^2 \rho$) for selected P--P and Se--Se atom pairs (see Fig.\ \protect\ref{fig1})  in three phases (LP ($R\bar{3}$), HP-I ($C2/m$) and HP-II ($P\bar{3}1m$)) of FePSe$_3$ at several pressures calculated using the B3LYP-13\% functional. $R$ is the interatomic distance, $R_1 + R_2$ is the sum of distances from the critical point to the atoms.  See text for details.  }
	\label{table4}     
	\centering 
	\renewcommand{\arraystretch}{0.8}
	\begin{tabular}{lllllll} 
		\\
		\hline 
		$P$ & Atom pair     & $R$    &  $\rho$   & $\bigtriangledown^2 \rho$   & $R_1 + R_2$  & CP type \\   
        (GPa) &       & (\AA)    &  (a.u.)   & (a.u.)   & (\AA)  &   \\       	      
		\hline
		\multicolumn{7}{c}{LP ($R\bar{3}$)} \\
		0 &P$_1$--P$_2$     & 2.182  &0.129	 & -0.219    &2.182       & (3,-1)    \\
   		  &P$_1$--P$_3$     & 5.952  &       &           &            &no bond   \\
		  &Se$_1$--Se$_2$   & 4.390  &0.003	 & 0.007     &4.396       & (3,-1)      \\
          &Se$_1$--Se$_3$   & 4.532  &0.002  & 0.007     &4.535       & (3,-1)     \\
          
		3 & P$_1$--P$_2$	& 2.173  &0.131	 &-0.227	 &2.173       & (3,-1)      \\
		  &P$_1$--P$_3$     & 5.444  &       &           &            &no bond       \\ 
		  
		6 & P$_1$--P$_2$	& 2.163  &0.134  &-0.237	 &2.162       & (3,-1)     \\
		  &P$_1$--P$_3$     & 5.196  &       &           &            &no bond       \\
		  
		\multicolumn{7}{c}{ } \\	
		\multicolumn{7}{c}{HP-I ($C2/m$)} \\
		6  & P$_1$--P$_2$   &2.165	 &0.134	 &-0.238	 &2.165	      & (3,-1)	 \\
 		   &P$_1$--P$_3$    &4.456	 &   	 &  	     &            &no bond  \\
 		   
		10 & P$_1$--P$_2$   &2.157	 &0.136  &-0.248	 &2.157       & (3,-1)  \\
		   &P$_1$--P$_3$    &4.268   &    	 &           &            &no bond   \\
		   & Se$_1$--Se$_2$ &3.483   &0.013	 &0.028      &3.485       &(3,-1) \\
		   &Se$_1$--Se$_3$  &3.450   &0.014  &0.029      &3.453       &(3,-1)      \\

		13 & P$_1$--P$_2$   &2.152	 &0.137	 &-0.254	 &2.152      & (3,-1)    	 \\
		   &P$_1$--P$_3$    &4.171	 &    	 &           &            &no bond   \\
		   
		\multicolumn{7}{c}{ } \\	
		\multicolumn{7}{c}{HP-II ($P\bar{3}1m$)} \\
		13&P$_1$--P$_2$     &3.047	&0.022	& 0.040   	&3.047        & (3,-1)   \\
		  & P$_1$--P$_3$	&2.293	&0.109	&-0.136 	&2.293   	  & (3,-1)   \\

		20&P$_1$--P$_2$     &2.541	&0.063	&0.010	    &2.541        & (3,-1)   \\
		  & P$_1$--P$_3$	&2.413	&0.089	&-0.061 	&2.413	      & (3,-1)   \\	
		  & Se$_1$--Se$_2$  & 2.830 &0.039	&0.074      &2.836        & (3,-1)   \\   
		  &Se$_1$--Se$_3$   &3.256  &0.020  &0.042      &3.260        & (3,-1)   \\
		  
		30& P$_1$--P$_2$	&2.369	&0.087	&-0.045	    &2.369	     & (3,-1)    \\
		  &P$_1$--P$_3$     &2.402	&0.090	&-0.062	    &2.402       & (3,-1)    \\
		\hline
	\end{tabular}
\end{table*}

\begin{table*}
	\footnotesize
	\caption{Several properties (the electron density $\rho$ and its Laplacian $\bigtriangledown^2 \rho$, the $|V(\textbf{r})| / G(\textbf{r})$ ratio, the total energy density $H(\textbf{r})$ (all in atomic units) and the kinetic energy per electron $G(\textbf{r})/\rho(\textbf{r})$)
		calculated for the selected P--P and Se--Se (see Fig.\ \protect\ref{fig1}) bond CPs of FePSe$_3$ at three pressures $P$ (in GPa) using the B3LYP-13\% functional. $R$ (in \AA) is the interatomic distance. See text for details. }
	\label{table5}     
	\centering 
	\renewcommand{\arraystretch}{0.8}
	\begin{tabular}{lllllccll} 
		\\
		\hline 
		$P$ & Atom pair     & $R$    &  $\rho$   & $\bigtriangledown^2 \rho$   &  $|V(\textbf{r})| / G(\textbf{r})$ & $H(\textbf{r})$  & $G(\textbf{r})/\rho(\textbf{r})$ & CP type \\   
		\hline
		\multicolumn{9}{c}{LP ($R\bar{3}$) } \\
		0 &P$_1$--P$_2$     & 2.182  &0.129	  & -0.219    &4.226   &-0.079  & 0.190  & (3,-1)    \\
		  &P$_1$--P$_3$     & 5.952  &        &      &        &     &  &  no bond   \\
	      &Se$_1$--Se$_2$   & 4.390  &0.003   & 0.007     &0.778   &0.000   &0.495  & (3,-1)      \\
		  &Se$_1$--Se$_3$   & 4.532  &0.002   & 0.007     &0.723   &0.000   &0.535  & (3,-1)     \\
	
		\multicolumn{9}{c}{ } \\	
		\multicolumn{9}{c}{HP-I ($C2/m$) } \\
		10 & P$_1$--P$_2$   &2.157  &0.136    &-0.248 &4.381 &-0.088  &0.191 & (3,-1)  \\
		   &P$_1$--P$_3$    &4.268   &         &         &      &       &     & no bond   \\
		   & Se$_1$--Se$_2$ &3.483  &0.013	  &0.028  &0.964 &0.000   &0.500 &(3,-1) \\
	       &Se$_1$--Se$_3$  &3.450  &0.014    &0.029  &0.961 &0.000   &0.500 &(3,-1)  \\
	          
		\multicolumn{9}{c}{ } \\	
		\multicolumn{9}{c}{HP-II ($P\bar{3}1m$) } \\
	  20&P$_1$--P$_2$       &2.541  &0.063	&0.010	&1.867   &-0.017     &0.223  & (3,-1)   \\
		  & P$_1$--P$_3$	&2.413  &0.089	&-0.061 &2.774   &-0.035     &0.308  & (3,-1)   \\	  
		  & Se$_1$--Se$_2$  &2.830 &0.039	&0.074  &1.115   &-0.002    &0.529  & (3,-1)   \\   
	   	  &Se$_1$--Se$_3$   &3.256 &0.020   &0.042  &1.058   &-0.001    &0.548  & (3,-1)   \\
		
		\hline
	\end{tabular}
\end{table*}

\begin{table*}
  	\footnotesize
   	\caption{The results of topological analysis (as in Table\ \protect\ref{table5}) for the PBE0 and M06 DFT functionals illustrating weak sensitivity of properties to the functional type.}
       	\label{table6}     
       	\centering 
       	\renewcommand{\arraystretch}{0.8}
       	\begin{tabular}{lllllccll} 
       		\\
       		\hline 
       		$P$ & Atom pair     & $R$    &  $\rho$   & $\bigtriangledown^2 \rho$   &  $|V(\textbf{r})| / G(\textbf{r})$ & $H(\textbf{r})$  & $G(\textbf{r})/\rho(\textbf{r})$ & CP type \\   
       		\hline

       		\multicolumn{9}{c}{ } \\
       		\multicolumn{9}{c}{LP ($R\bar{3}$) PBE0} \\
       	0   &P$_1$--P$_2$   & 2.150  &0.137	  & -0.260    &4.584   &-0.090  & 0.183  & (3,-1)    \\
       		&P$_1$--P$_3$     & 5.711  &        &      &        &     &  &  no bond   \\
       		&Se$_1$--Se$_2$   & 4.127  &0.005   & 0.010     &0.876   &0.000   &0.468  & (3,-1)      \\
       		&Se$_1$--Se$_3$   & 4.261  &0.004   & 0.009     &0.842   &0.000   &0.490  & (3,-1)     \\		
       		
       		\multicolumn{9}{c}{ } \\
       		\multicolumn{9}{c}{LP ($R\bar{3}$) M06} \\
       	0   &P$_1$--P$_2$   & 2.157  &0.134	& -0.240    &4.333  &-0.086  & 0.192  & (3,-1)    \\
       		&P$_1$--P$_3$     & 5.579  &        &      &        &     &  &  no bond   \\
       		&Se$_1$--Se$_2$   & 3.985  &0.006   & 0.012     &0.926   &0.000   &0.466  & (3,-1)      \\
       		&Se$_1$--Se$_3$   & 4.130  &0.005   & 0.011     &0.897   &0.000   &0.483  & (3,-1)     \\		

       		\multicolumn{9}{c}{ } \\	
       		\multicolumn{9}{c}{HP-I ($C2/m$) PBE0} \\
       	10  & P$_1$--P$_2$ &2.130  &0.143    &-0.286 &4.704 &-0.098  &0.185 & (3,-1)  \\
       		&P$_1$--P$_3$     &4.210   &         &         &      &       &     & no bond   \\
       		& Se$_1$--Se$_2$  &3.382  &0.016    &0.033  &0.975 &0.000   &0.512 &(3,-1) \\
       		&Se$_1$--Se$_3$   &3.405  &0.015    &0.032  &0.976 &0.000   &0.513 &(3,-1)  \\
       	
       		\multicolumn{9}{c}{ } \\	
       		\multicolumn{9}{c}{HP-I ($C2/m$) M06} \\
       	10  & P$_1$--P$_2$ &2.130  &0.141    &-0.272 &4.506 &-0.095  &0.193 & (3,-1)  \\
       		&P$_1$--P$_3$     &4.293   &        &         &      &       &     & no bond   \\
       		& Se$_1$--Se$_2$  &3.456  &0.014    &0.030  &0.945 &0.000   &0.516 &(3,-1) \\
       		&Se$_1$--Se$_3$   &3.492  &0.013    &0.028  &0.947 &0.000   &0.515 &(3,-1)  \\
       		       		
       		\multicolumn{9}{c}{ } \\	
       		\multicolumn{9}{c}{HP-II ($P\bar{3}1m$) PBE0} \\
       	20  &P$_1$--P$_2$   &2.451 &0.074	&-0.014	&2.165   &-0.025     &0.296  & (3,-1)   \\
       		& P$_1$--P$_3$	  &2.383 &0.095	&-0.082 &2.970   &-0.041     &0.222  & (3,-1)   \\		
       		& Se$_1$--Se$_2$  &2.754 &0.045	&0.078  &1.191   &-0.005    &0.534  & (3,-1)   \\   
       		&Se$_1$--Se$_3$   &3.162 &0.024 &0.049	&1.105   &-0.001    &0.559  & (3,-1)   \\	   	  
       		
       		\multicolumn{9}{c}{ } \\	
       		\multicolumn{9}{c}{HP-II ($P\bar{3}1m$) M06} \\
       	20  &P$_1$--P$_2$   &2.480  &0.070	&-0.002	&2.025   &-0.022     &0.310  & (3,-1)   \\
       		& P$_1$--P$_3$	  &2.449  &0.082	&-0.041 &2.512   &-0.031     &0.245  & (3,-1)   \\	
       		& Se$_1$--Se$_2$  &2.829 &0.039  	&0.077  &1.107   &-0.002    &0.554  & (3,-1)   \\   
       		&Se$_1$--Se$_3$   &3.227 &0.021	    &0.046  &1.059   &-0.001    &0.571  & (3,-1)   \\	   	  
       		\hline
       	\end{tabular}
       \end{table*}

\end{document}